\documentclass[%
 reprint,
superscriptaddress,
 amsmath,amssymb,
 aps,
]{revtex4-1}

\usepackage{graphicx}
\usepackage{dcolumn}
\usepackage{bm}
\usepackage{booktabs}
\usepackage{longtable}

\def\hyph{-\penalty0\hskip0pt\relax}

\def\Nepg{$^{22}\text{Ne}(p,\gamma)^{23}\text{Na}$}

\begin{document}

\preprint{APS/123-QED}

\title{First inverse kinematics study of the $^{22}$Ne$(p,\gamma)^{23}$Na reaction and its role in AGB star and classical nova nucleosynthesis}

\author{M. Williams}
 \email{mwilliams@triumf.ca}
\affiliation{Department of Physics, University of York, Heslington, York, UK, YO10 5DD}
\affiliation{TRIUMF, 4004 Wesbrook Mall, Vancouver, BC, Canada, V6T 2A3}
\author{A. Lennarz}
\affiliation{TRIUMF, 4004 Wesbrook Mall, Vancouver, BC, Canada, V6T 2A3}
\author{A. M. Laird}
\affiliation{Department of Physics, University of York, Heslington, York, UK, YO10 5DD}
\affiliation{The NuGrid collaboration, http://www.nugridstars.org}
\author{U. Battino}
\affiliation{University of Edinburgh, School of Physics and Astrophysics, Edinburgh EH9 3FD, UK}
\affiliation{The NuGrid collaboration, http://www.nugridstars.org}
\author{J. Jos\'e}
\affiliation{Departament de F\'isica, Universitat Polit\`ecnica de Catalunya \& Institut d'Estudis Espacials de Catalunya (IEEC), C. Eduard Maristany 16, E-08019 \& Ed. Nexus-201, C. Gran Capit\`a, 2-4, E-08034, Barcelona, Spain}
\author{D. Connolly}
\altaffiliation[Present address: ]{Los Alamos National Laboratory, Los Alamos, New Mexico 87545, USA}
\affiliation{TRIUMF, 4004 Wesbrook Mall, Vancouver, BC, Canada, V6T 2A3}
\author{C. Ruiz}
\affiliation{TRIUMF, 4004 Wesbrook Mall, Vancouver, BC, Canada, V6T 2A3}
\author{A. Chen}
\affiliation{Department of Physics and Astronomy, McMaster University, Hamilton, ON, Canada, L8S 4L8}
\author{B. Davids}
\affiliation{TRIUMF, 4004 Wesbrook Mall, Vancouver, BC, Canada, V6T 2A3}
\affiliation{Department of Physics, Simon Fraser University, 8888 University Drive, Burnaby, BC, V5A 1S6, Canada}
\author{N. Esker}
\altaffiliation[Present address: ]{San Jos\'e State University, 1 Washington Square,
Duncan Hall 518 San Jos\'e, CA 95192-0101, USA}
\affiliation{TRIUMF, 4004 Wesbrook Mall, Vancouver, BC, Canada, V6T 2A3}
\author{B. R. Fulton}
\affiliation{Department of Physics, University of York, Heslington, York, UK, YO10 5DD}
\author{R. Garg}
\altaffiliation[Present address: ]{University of Edinburgh, School of Physics and Astrophysics, Edinburgh EH9 3FD, UK}
\affiliation{Department of Physics, University of York, Heslington, York, UK, YO10 5DD}
\author{M. Gay}
\affiliation{Columbia University, 116th St \& Broadway, New York, NY 10027, USA}
\author{U. Greife}
\affiliation{Colorado School of Mines, Golden, CO, USA}
\author{U. Hager}
\affiliation{National Superconducting Cyclotron Laboratory, Michigan State University, East Lansing, MI  48824, USA}
\author{D. Hutcheon}
\affiliation{TRIUMF, 4004 Wesbrook Mall, Vancouver, BC, Canada, V6T 2A3}
\author{M. Lovely}
\affiliation{Colorado School of Mines, Golden, CO, USA}
\author{S. Lyons}
\affiliation{National Superconducting Cyclotron Laboratory, Michigan State University, East Lansing, MI  48824, USA}
\affiliation{The Joint Institute for Nuclear Astrophysics--Center for the Evolution of the Elements, Michigan State University, East Lansing, Michigan 48824, USA}
\author{A. Psaltis}
\affiliation{Department of Physics and Astronomy, McMaster University, Hamilton, ON, Canada, L8S 4L8}
\author{J. E. Riley}
\affiliation{Department of Physics, University of York, Heslington, York, UK, YO10 5DD}
\author{A. Tattersall}
\affiliation{University of Edinburgh, School of Physics and Astrophysics, Edinburgh EH9 3FD, UK}
\affiliation{The NuGrid collaboration, http://www.nugridstars.org}




\date{\today}

\begin{abstract}
\begin{description}
\item[Background]
Globular clusters are known to exhibit anomalous abundance trends such as the sodium-oxygen anti-correlation. This trend is thought to arise via pollution of the cluster interstellar medium from a previous generation of stars. Intermediate-mass asymptotic giant branch stars undergoing Hot Bottom Burning (HBB) are a prime candidate for producing sodium-rich oxygen-poor material, and then expelling this material via strong stellar winds. The amount of $^{23}$Na produced in this environment has been shown to be sensitive to uncertainties in the $^{22}$Ne$(p,\gamma)^{23}$Na reaction rate. The $^{22}$Ne$(p,\gamma)^{23}$Na reaction is also activated in classical nova nucleosynthesis, strongly influencing predicted isotopic abundance ratios in the Na-Al region. Therefore, improved nuclear physics uncertainties for this reaction rate are of critical importance for the identification and classification of pre-solar grains produced by classical novae.
\item[Purpose]
At temperatures relevant for both HBB in AGB stars and classical nova nucleosynthesis, the $^{22}$Ne$(p,\gamma)^{23}$Na reaction rate is dominated by narrow resonances, with additional contribution from direct capture. This study presents new strength values for seven resonances, as well as a study of direct capture.
\item[Method]
The experiment was performed in inverse kinematics by impinging an intense isotopically pure beam of $^{22}$Ne onto a windowless H$_{2}$ gas target. The $^{23}$Na recoils and prompt $\gamma$ rays were detected in coincidence using a recoil mass separator coupled to a $4\pi$ bismuth-germanate (BGO) scintillator array surrounding the target.
\item[Results]
For the low energy resonances, located at center of mass energies of 149, 181 and 248 keV, we recover stength values of $\omega\gamma_{149} = 0.17^{+0.05}_{-0.04}$, $\omega\gamma_{181} = 2.2 \pm 0.4$, and $\omega\gamma_{248} = 8.2 \pm 0.7$ $\mu$eV, respectively. These results are in broad agreement with recent studies performed by the LUNA and TUNL groups. However, for the important reference resonance at 458 keV we obtain a strength value of $\omega\gamma_{458} = 0.44 \pm 0.02$ eV, which is significantly lower than recently reported values. This is the first time that this resonance has been studied completely independently from other resonance strengths. In the case of direct capture, we recover an S-factor of 60 keV$\cdot$b, consistent with prior forward kinematics experiments.
\item[Conclusions]
In summary, we have performed the first direct measurement of $^{22}$Ne$(p,\gamma)^{23}$Na in inverse kinematics. Our results are in broad agreement with the literature, with the notable exception of the 458 keV resonance, for which we obtain a lower strength value. We assessed the impact of the present reaction rate in reference to a variety of astrophysical environments, including AGB stars and classical novae. Production of $^{23}$Na in AGB stars is minimally influenced by the factor of 4 increase in the present rate compared to the STARLIB-2013 compilation. The present rate does however impact upon the production of nuclei in the Ne-Al region for classical novae, with dramatically improved uncertainties in the predicted isotopic abundances present in the novae ejecta.

\end{description}
\end{abstract}

\pacs{Valid PACS appear here}
\maketitle


\section{\label{sec:intro}Introduction}

Globular clusters (GCs) are dense associations of stars that formed in the early universe. Containing some of the oldest observed stars, these remarkable objects provide estimates for the age of our galaxy as well as a lower limit on the age of the universe \cite{Krauss65}. In addition to their cosmological importance, GCs are important test sites for the study of galactic chemical evolution as they are thought to consist of a single coeval population of stars. However, advances in optical astronomy have challenged this simple picture, with many globular clusters containing multiple generations of stars accompanied by anomalous abundance correlations \cite{gratton2004,gratton2012,carretta2011,yong2003}. One such abundance trend is the sodium-oxygen anti-correlation, which is observed ubiquitously over all well-studied globular clusters to date. This abundance trend is not reproduced in field stars however, suggesting that the cluster environment itself has a profound influence.


The site responsible for the Na-O anti-correlation must reach temperatures sufficient for activation of both the CNO and NeNa cycles. However, this abundance trend is observed in many stars that could not have reached the required core temperatures for nucleosynthesis beyond $A=20$ \cite{prantzos2007}. This leads to the idea that the cluster environment must have been enriched by a previous generation of stars. Massive ($M \geqslant 4 M_{\odot}$) AGB stars undergoing Hot Bottom Burning (HBB) have been put forward as prime candidates for polluting the cluster interstellar medium (ISM) \cite{antona1982,ventura2008}. Other potential scenarios could also contribute, such as: fast rotating massive stars \cite{Descressin2007}, massive binaries \cite{Mink2009}, and supermassive $(M \approx 10^{4} M_{\odot})$ stars \cite{denissenkov2014}; though AGB stars remain the most likely site to be the dominant source of sodium-rich oxygen-poor material \cite{Renzini2008,lee2010}. Here, $^{23}$Na is produced at the base of the convective hydrogen envelope by radiative proton capture on $^{22}$Ne; the third most abundant nuclide produced in core helium burning \cite{Buch2006}. According to stellar evolution calculations \cite{Vent2005}, temperatures at the base of the convective envelope reach to approximately 0.1 GK which is sufficient to activate the NeNa and MgAl burning cycles. This leads to a rise in the Na and Al content of the surrounding stellar envelope as the processed material is brought to the surface by successive third dredge up (TDU) episodes as the star undergoes thermal pulses. The oxygen content is simultaneously reduced by activation of the ON cycle, resulting in the observed NaO anti-correlation.      

The $^{22}$Ne$(p,\gamma)^{23}$Na reaction also plays a role in classical novae nucleosynthesis. A sensitivity study performed by Iliadis \textit{et al.} \cite{iliadis-AJS142-2002} showed that in the case of oxygen-neon (ONe) novae with underlying white dwarf masses of 1.15 and 1.25 $M_{\odot}$, reaching respective peak temperatures of $T_{peak} = 0.231$ and 0.251 GK, the final abundance of $^{22}$Ne was altered by up to 6 orders of magnitude as a result of varying the rate within its upper and lower uncertainty limits. Whereas, in the case of carbon-oxygen (CO) novae with a 1 $M_{\odot}$ white dwarf mass ($T_{peak} = 0.17$ GK), $^{22}$Ne was affected by a factor of 100, $^{23}$Na by a factor of 7, $^{24}$Mg by a factor of 5, as well as factor of 2 changes in $^{20}$Ne, $^{21}$Ne, $^{25}$Mg, $^{26}$Mg, $^{26}$Al and $^{27}$Al.  

The $^{22}$Ne$(p,\gamma)^{23}$Na reaction rate has carried an exceptionally large uncertainty due to a number of (until recently) unobserved resonances, many of which reside in the Gamow window for both classical novae and HBB in AGB stars. The discrepancy in available rate compilations spans a factor of 1000 between the NACRE \cite{ANGULO19993} and STARLIB-2013 \cite{sallaska2013} compilations. This situation was recently changed by an experiment performed at the LUNA facility \cite{LUNA2010}, in which the strengths of three new resonances at $E_{c.m.} = 149, 181,$ and $248$ keV (where $E_{c.m.}$ is the resonance energy in the center of mass frame) were measured by Cavanna \textit{et al.} \cite{Cavanna-PRL115-2015}. The existence of the two lowest energy resonances were subsequently confirmed by Kelly \textit{et al.} \cite{Kelly-PRC95-2017} in a study performed at the LENA facility \cite{CESARATTO2010,LONGLAND2006452}. This latter study measured the strengths of the aforementioned resonances relative to that of the 458 keV resonance reported in Ref \cite{Kelly-PRC92-2015}. The LUNA study by Cavanna \textit{et al.} also included direct upper limits for possible resonances at $E_{cm} = 68$ and 100 keV. These resonances were tentatively reported in a ($^{3}$He,$d$) transfer study by Powers \textit{et al.} \cite{Powers71}, but could not be confirmed in a later study by Hale \textit{et al.} \cite{Hale2002}. Moreover, the corresponding states in $^{23}$Na at $E_{x} =$ 8894 and 8862 keV were not observed in a $^{23}$Na$(p,p\prime)^{23}$Na measurement by Moss \textit{et al.} \cite{moss1976}, nor were they seen in a more recent spectroscopic study by Jenkins \textit{et al.} \cite{Jenkins-PRC87-2013} using Gammasphere. These resonances have thus not been considered for both the reaction rates put forward by Kelly \textit{et al.} and the present work. It is perhaps unsurprising that a subsequent attempt by the LUNA collaboration to measure these resonances directly, using a $\gamma$-ray spectrometer comprised of BGO instead of HPGe detectors, could not positively identify any yield from these resonances \cite{Ferraro-PRL-2018}. Although their newly obtained upper limits effectively remove the 100 keV resonance from contention as a significant contributor to the  $^{22}$Ne$(p,\gamma)^{23}$Na reaction rate, the 68 keV resonance remains a potential contributor, thus defining the upper limit of the new LUNA rate at temperatures below 0.1 GK.

The present work reports on the first inverse kinematics study of the $^{22}$Ne$(p,\gamma)^{23}$Na reaction rate, performed using the Detector of Recoils And Gamma-rays Of Nuclear reactions (DRAGON). Here we present strength measurements for the three low energy resonances at center of mass energies of 149, 181 and 248 keV, along with the important reference resonances at 458, 610, 632 and 1222 keV (center of mass). The non-resonant cross section was also measured in the energy range of $282 \leqslant E_{c.m.} \leqslant 511$ keV. All previous measurements of the $^{22}$Ne$(p,\gamma)^{23}$Na reaction have been carried out in forward kinematics. The present study is thereby subject to a different set of systematic uncertainties than those already found in the literature. It is important, particularly in the case of reference resonances, to derive consistent strength values and S-factors (in the case of direct capture) from a variety of experimental techniques.


\section{\label{sec:exp}Experiment Description}


This study was performed using the Detector of Recoils And Gammas Of Nuclear reactions (DRAGON) \cite{Hutcheon-NIMPRSA498-2003}, located in the ISAC-I experimental hall \cite{laxdal2003} at TRIUMF, Canada's national laboratory for particle and nuclear physics. An isotopically pure beam of $^{22}\mathrm{Ne}$ was generated by the Multi Charge Ion Source (MCIS) \cite{Jayamanna2010} in the $q=4^{+}$ charge state, which was then accelerated to lab energies in the range of $E_{lab} = 161-1274$ keV/$u$ via the ISAC-I Radio-Frequency Quadrupole (RFQ) and Drift-Tube Linac (DTL). The beam was delivered to the DRAGON experiment area with a maximum intensity of $5 \times 10^{12}$ pps, and FWHM beam energy spread of $\Delta E/E \leqslant 0.4 \%$. 


The DRAGON facility consists of three primary components: (1) a windowless differentially pumped gas target surrounded by a $4\pi$ $\gamma$-ray detector array, (2) an electromagnetic vacuum-mode mass separator, and (3) a series of heavy ion detectors located at the focal plane of the separator. The ion-optical configuration of the separator consists of two pairs of magnetic and electric dipole field elements, interspersed with quadrupole and sextupole lenses, as well as strategically placed slit systems for increased beam suppression. 

The DRAGON $\gamma$-ray detector array, which surrounds the gas target, is comprised of 30 BGO scintillator crystals and photo-multiplier tubes (PMTs). The close-packed geometry of the array around the gas target vacuum box gives a total solid-angle coverage of 92\%. The heavy-ion detectors employed for this study were a pair Micro-Channel Plate (MCP) detectors, followed by a Double-sided Silicon Strip Detector (DSSD) \cite{tengblad2004}. The pair of MCP detectors form a local transmission time-of-flight (TOF) measurement system, whereby ions can be identified via their transit time across a small section of beam-line. The transmitted ions are then stopped in the DSSD, where their kinetic energy is measured. Coincidences between recoils and prompt $\gamma$-rays were identified by a timestamp-based algorithm \cite{Greg2014}. 

\begin{figure}
 \includegraphics[width=0.45\textwidth]{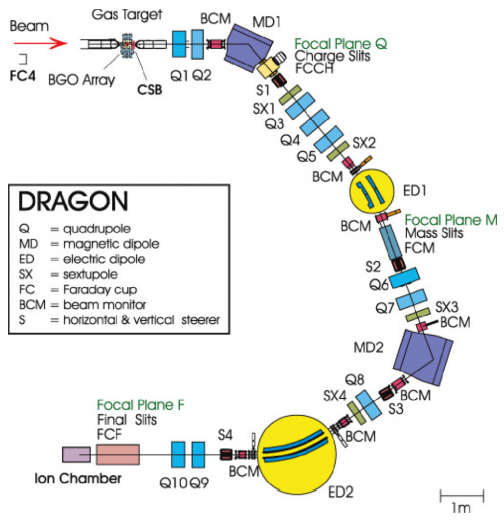}
\caption{Schematic of the DRAGON recoil separator. The electromagnetic elements, slit positions, and Faraday cups are labeled.} \label{fig:dragon}
 \end{figure}





The present experiment has several advantages over the techniques utilized in already published works for this reaction. Difficulties relating to the gaseous nature of both reactant species, such as contaminating background reactions and uncertain target stoichiometry, are circumvented by conducting the experiment in inverse kinematics with a window-less recirculated gas target. The stopping power of the beam through the target is also directly measured by tuning beam through DRAGON's first magnetic dipole (see section \ref{sec:beam_energy}).

\section{Data Analysis}

A total of sixteen successful yield measurements were made, at fourteen different beam energies. The present work targets seven resonances at center of mass energies of: 1222, 632, 610, 458, 248, 181, and 149 keV. The strength of the 632-keV resonance was measured at three different target pressures, in order to exclude contamination from the 610-keV resonance. The non-resonant cross section was also measured at seven different beam energies in the center of mass energy range from 282 keV to 511 keV.

\subsection{\label{yield}Thick target yield, reaction cross section and resonance strength}

Laboratory experiments of nuclear reaction cross sections (and resonance strengths) measure the reaction \textit{yield}, which is defined per incident beam ion as:

\begin{equation} \label{eqn:yield1}
	Y = \frac{N^{tot}_{r}}{N_{b}}
\end{equation}

where $N^{tot}_{r}$ is the total number of reactions that occur, and $N_{b}$ is the number of beam ions incident on the target. At DRAGON, the total number of reactions is inferred by combining the number of detected recoils with the systematics of the experiment. Therefore, equation \ref{eqn:yield1} can be re-written as:

\begin{equation} \label{eqn:yield2}
	Y = \frac{N^{det}_{r}}{N_{b} \, \varepsilon_{\mathrm{DRA}}}
\end{equation}

where $\varepsilon_{\mathrm{DRA}}$ is the product of all efficiencies affecting the number of detected recoils, $N^{\mathrm{det}}_{r}$. Recoils can be measured either in coincidence with a $\gamma$-ray hit in the BGO detectors or without a detected $\gamma$-ray, referred to as coincidence and singles events respectively. The systematics of the two aforementioned event designation are slightly different, with the former influenced by the detection efficiency of the BGO array. The total detection efficiencies pertaining to singles and coincidence events are given as:

\begin{equation} \label{eqn:singles_eff}
\varepsilon_{\textrm{DRA}}^{\mathrm{sing}} = f_{q} \cdot \tau_{\mathrm{MCP}} \cdot \varepsilon_{\mathrm{MCP}} \cdot \varepsilon_{\mathrm{DSSD}} \cdot \tau_{\mathrm{rec}} \cdot \lambda_{\mathrm{tail}}
\end{equation}

\begin{equation} \label{eqn:coinc_eff}
\varepsilon_{\textrm{DRA}}^{\mathrm{coinc}} = f_{q} \cdot \tau_{\mathrm{MCP}} \cdot \varepsilon_{\mathrm{MCP}} \cdot \varepsilon_{\mathrm{DSSD}} \cdot \varepsilon_{\gamma} \cdot \lambda_{\mathrm{coinc}}
\end{equation}

The first four terms in Equations \ref{eqn:singles_eff} and \ref{eqn:coinc_eff} are common to both singles and coincidence events. These are: the recoil charge state fraction $(f_{q})$, MCP transmission efficiency $(\tau_{\mathrm{MCP}})$, MCP detection efficiency $(\epsilon_{MCP})$ and detection efficiency of the DSSD $(\varepsilon_{\mathrm{DSSD}})$. $\lambda_{\mathrm{tail}}$ is the live time fraction of the focal plane DAQ, whereas $\lambda_{\mathrm{coinc}}$ is the live time fraction where both the target (head) and focal plane (tail) DAQs are able to accept new triggers \cite{Greg2014}.

The recoil transmission efficiency, $\tau_{\mathrm{rec}}$, relates to the number of recoils that are produced within the acceptances of the separator. Obtained through simulation, this quantity depends on the kinematics of the radiative capture reaction and its effect on the transmission of recoils through the separator. The recoil-gamma coincidence efficiency $(\varepsilon_{\gamma})$ is the probability that a transmitted recoil will be recorded in coincidence with a prompt $\gamma$-ray detected by the BGO array. This quantity is also obtained via simulation, calculated as:

\begin{equation}
\varepsilon_{\gamma} = \frac{N^{\mathrm{sim}}_{\mathrm{coinc}}}{N^{\mathrm{sim}}_{\mathrm{react}}},
\end{equation}

where $N^{\mathrm{sim}}_{\mathrm{react}}$ is the simulated number of reactions, and $N^{\mathrm{sim}}_{\mathrm{coinc}}$ is the total number of $\gamma$-rays detected in coincidence with a recoil transmitted to the focal plane. Note that this definition of the recoil-$\gamma$ coincidence efficiency already accounts for the transmission of recoils, therefore, $\tau_{\mathrm{rec}}$ need not be included in the total coincidence efficiency.

The total yield is related to the reaction cross section, integrated over the entire target length, by:

\begin{equation}
Y = \sigma \, n_{t} \, L_{\mathrm{eff}},
\end{equation}

where $\sigma$ is the total reaction cross section, $L_{\mathrm{eff}}$ is the effective target length, and $n_{t}$ is the number density of the hydrogen gas target. The number density is determined from the average pressure and temperature of the target via the ideal gas law. 

The reaction cross section can be used to derive the \textit{astrophysical S-factor}, $S(E)$, via the following definition:

\begin{equation}
    \sigma(E) \equiv \frac{1}{E} e^{-2\pi\eta} S(E),
\end{equation}

where $E$ is the center of mass energy and the term $e^{2\pi\eta}$ is the Gamow factor, which accounts for the $s$-wave penetrability at energies well-below the Coulomb barrier. This definition of the S-factor removes strongly energy dependent effects impacting the reaction cross section. 
For narrow resonances, wherein the resonance width is small compared to the target width, the reaction yield becomes the thick target yield $(Y \rightarrow Y_{\infty})$. With center of mass target thicknesses in the range of $7$ - $20$ keV, all the resonances considered in this study are sufficiently narrow to satisfy the thick target yield condition. For a narrow Breit-Wigner resonance the thick target yield is related to the resonance strength by:

\begin{equation}
\omega \gamma = \frac{2Y_{\infty}}{\lambda_{r}^{2}} \frac{m_{p}}{m_{p}+m_{t}} \, \epsilon_{\mathrm{lab}},
\end{equation}

where $\omega\gamma$ is the resonance strength in $eV$, $m_{p}$ and $m_{t}$ are the projectile and target masses (in $u$) respectively, $\epsilon_{\mathrm{lab}}$ is the laboratory frame stopping power (eV/cm$^{2}$), and $\lambda_{r}$ is the de Broglie wavelength (cm) associated with the relative energy of the resonance in the center of mass frame.



\subsection{\label{sec:beam_energy}Beam energy and stopping power}

The incident beam energy was measured by tuning through the first magnetic dipole (MD1) onto a downstream pair of slits. The slit plates are electrically isolated so as to enable current to be measured on each plate. The slit plates are unsuppressed however, and therefore do not permit measurement of absolute current. Nonetheless, with the slits closed to 2 mm, they do serve as accurate beam tuning diagnostics to center a given charge state through MD1. The beam energy is related to the MD1 field, as measured by its NMR probe, through: 

\begin{equation} \label{eqn:energy}
E/A = c_{\textrm{mag}} (qB/A)^{2} - \frac{1}{2uc^{2}}(E/A)^{2},
\end{equation}

Where $A$ is the atomic mass of the beam, $q$ is the beam charge state after the target, $B$ is the MD1 field (in Tesla) measured by its NMR probe, $u$ is the atomic mass unit, $c$ is the the speed of light, and $c_{\textrm{mag}}=48.15 \pm 0.07 \; \textrm{MeV} \; \textrm{T}^{2}$ is a constant related to the effective bending radius of MD1 \cite{Hutcheon-NIMPRA-2012}. The final term is a relativistic correction that has only a minor influence on the measured energy and is often neglected. 

The total energy lost across the gas target was measured by using Equation \ref{eqn:energy} to determine the beam energy with and without gas present in the target. In instances where the incident beam exceeds the rigidity limit of MD1, as was the case for the $E_{c.m.}=1222 \; \textrm{MeV}$ yield measurement, the outgoing beam energy is measured at several gas target pressures. The incident energy is then found by a linear extrapolation of the measured beam energies to zero-pressure. The stopping power across the target can be directly obtained by combining the measured energy loss and target number density. The ability to directly measure stopping powers in the lab is a key advantage of the DRAGON facility as systematic uncertainties related to the use of semi-empirical codes such as \textsc{SRIM} \cite{ZIEGLER2010} are avoided.

\subsection{\label{sec:beamnorm}Beam Normalization}

The total number of incident beam ions was determined by taking hourly beam current measurements using a Faraday cup (FC4) positioned approximately 2 m upstream of the gas target. Beam fluctuations within each data taking run were accounted for by relating these regular current measurements to the number target atoms scattered into two ion implanted silicon (IIS) detectors, mounted at $30^{\circ}$ and $57^{\circ}$ relative to the beam axis. The beam normalization coefficient, R, for a given run, is obtained as:

\begin{equation} \label{eqn:rfactor}
R = \frac{I}{e q} \frac{\Delta t \, P}{N_{p}} \, \varepsilon_{t}
\end{equation}

where $I$ is the beam current as measured by FC4 and $e  q$ is the charge of the incident beam ions. $\Delta t$ is a short time interval, immediately proceeding a Faraday cup reading, over which the target pressure $P$ and number of elastically scattered protons $N_{p}$ is measured. The beam transmission efficiency $(\varepsilon_{t})$ through the target apertures is measured after each re-tune of the beam by recording the ratio of current measured by FC1 (immediately downstream of the target) over the current measured by FC4. The average normalization coefficient over all runs within a given yield measurement, $\langle R\rangle$, can then be used with Equation \ref{eqn:nbeam} to determine the total number of beam ions:

\begin{equation} \label{eqn:nbeam}
N_{b} = \frac{\langle R\rangle N_{p}^{\textrm{tot}}}{\langle P \rangle},
\end{equation}

where $N_{p}^{\textrm{tot}}$ is now the total number of elastically scattered protons, and $\langle P \rangle$ is the average pressure measured over all runs.

\subsection{\label{sec:csd} $^{23}\textrm{Na}$ Charge State Distribution}

DRAGON is designed to accept only a single charge state through the separator to the focal plane detectors. Therefore, in order to recover the full reaction yield, the charge state fraction of the recoils to which DRAGON is tuned to accept must be known. For the present work, a stable beam of $^{23}$Na was tuned to DRAGON in order to measure the recoil charge state distributions. The incident beam energies and gas-target pressures were selected such that the outgoing beam would closely match the energies of the $^{23}$Na recoils from the targeted $^{22}$Ne$(p,\gamma)^{23}$Na yield measurements. 

\begin{figure}[h!]
 \includegraphics[width=0.45\textwidth]{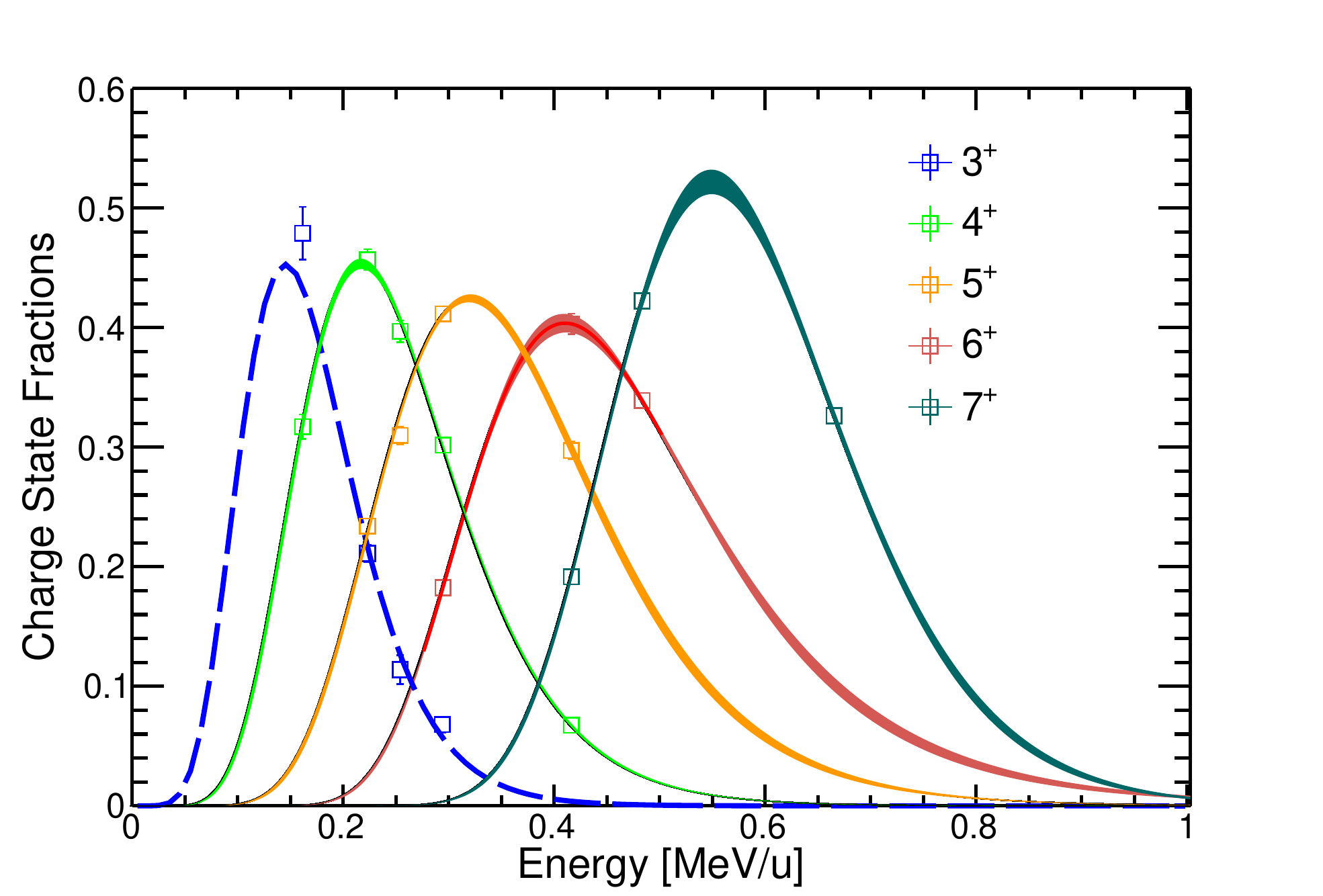}
\caption{Normalized charge state fractions for each charge state as a function of outgoing $^{23}$Na energy. The distributions are fit with the semi-empirical formula of Liu \textit{et al.} \cite{wenji}, with the shaded regions indicating the $1\sigma$ confidence limits of the fits. The $3^{+}$ fit did not converge due to a lack of data points on the rising portion of the distribution. Instead, the $q=3^{+}$ recoil charge state fraction, utilized for the $E_{c.m.} = 149$ keV yield measurement, was determined after the experiment at the outgoing recoil energy. The dashed blue curve is simply to guide the eye.} \label{fig:csd}
 \end{figure}

The charge state distributions were measured by tuning various charge states of the $^{23}$Na beam through the first magnetic dipole (MD1) with H$_{2}$ gas present in the target. The charge states are centered onto a Faraday cup (FCCH) positioned at the charge focal plane immediately downstream of MD1 (see schematic of DRAGON shown in Figure \ref{fig:dragon}). The resulting charge state distributions are then fit with a Gaussian normalized to unity. As a second step, the fraction of recoils in a given charge state as a function of outgoing $^{23}$Na energy were then fitted using the semi-empirical formula of Lui \textit{et al.} \cite{wenji}. The fit functions, and associated $1\sigma$ confidence bounds, were then evaluated for the outgoing recoil energies. The recoil charge state fractions for the lowest and highest energy measurements, at $E_{r} = 149$ and 1222 keV respectively, required special consideration. In the case of the $E_{r} = 149$ keV resonance the full charge state distribution could not be measured as $^{23}$Na ions emerging from the gas target in the $q = 2^{+}$ charge state could not be bent by MD1. Instead, the charge state distribution measurement was performed after the experiment with the gas target pressure set such that the outgoing $^{23}$Na ions would have the same energy as those DRAGON was tuned to accept during the $^{22}$Ne$(p,\gamma)^{23}$Na experimental run. The same procedure was utilized for determining the $q=9^{+}$ charge state fraction for the $E_{r} = 1222$ keV resonance since the charge state fractions were only measured at the outgoing $^{22}$Ne$(p,\gamma)^{23}$Na recoil energy, and were not measured over a large enough energy range so as to provide a good fit using a semi-empirical formula. 


\section{\label{sec:results}Results}

\begin{figure*}[httt]
\minipage{0.33\textwidth}
  \includegraphics[width=1.1\textwidth]{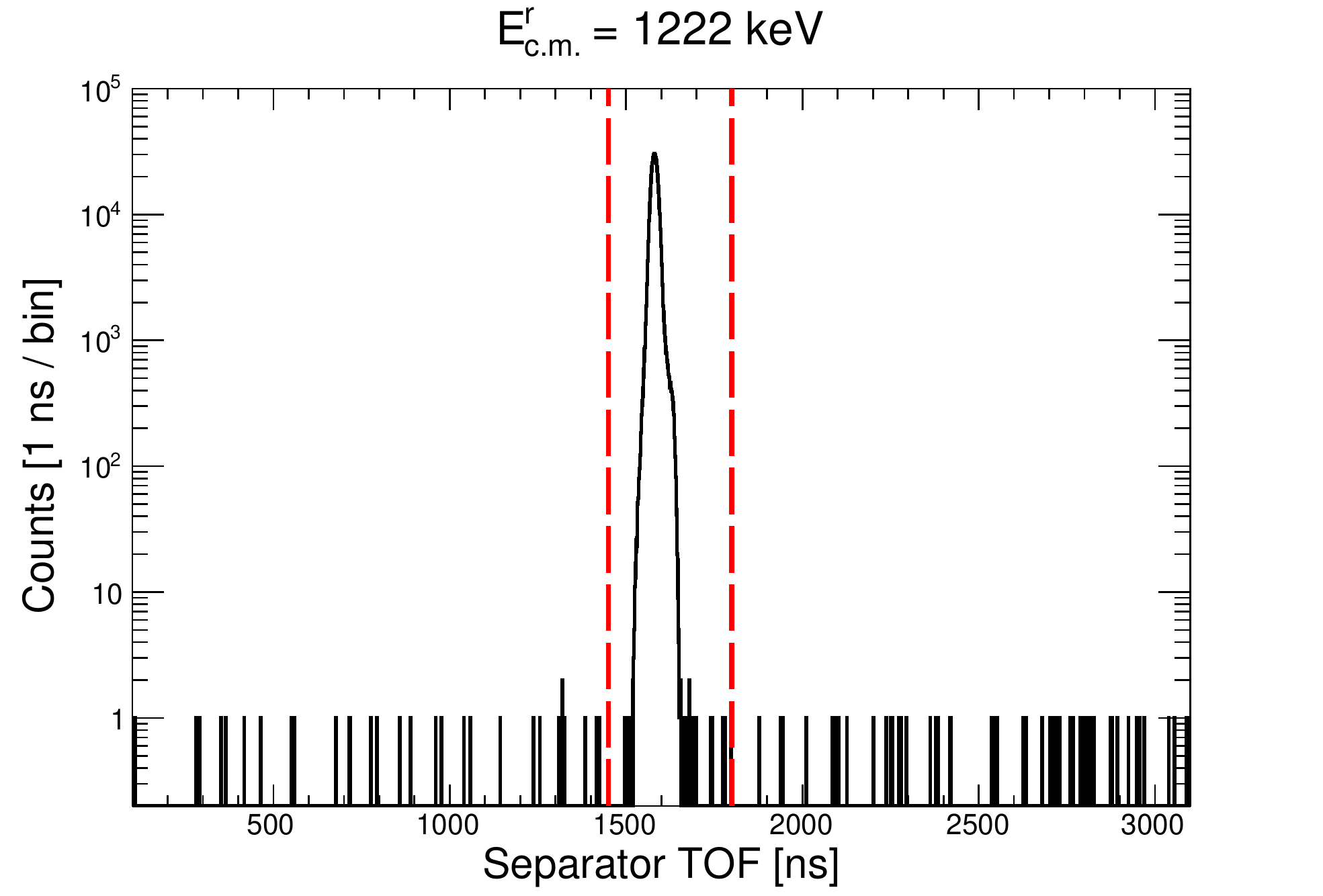}
\endminipage\hfill
\minipage{0.33\textwidth}
  \includegraphics[width=1.1\textwidth]{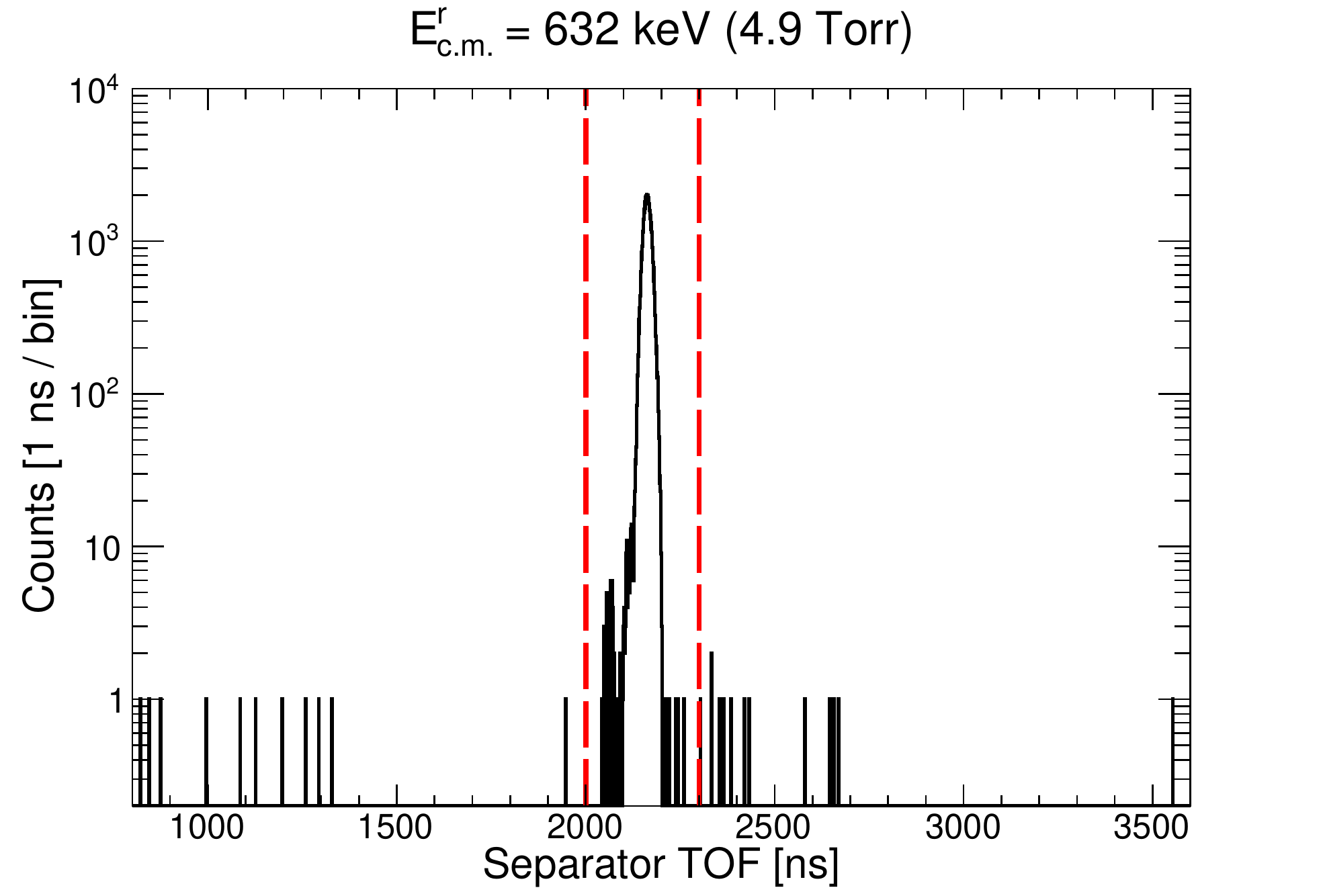}
\endminipage\hfill
\minipage{0.33\textwidth}
  \includegraphics[width=1.1\textwidth]{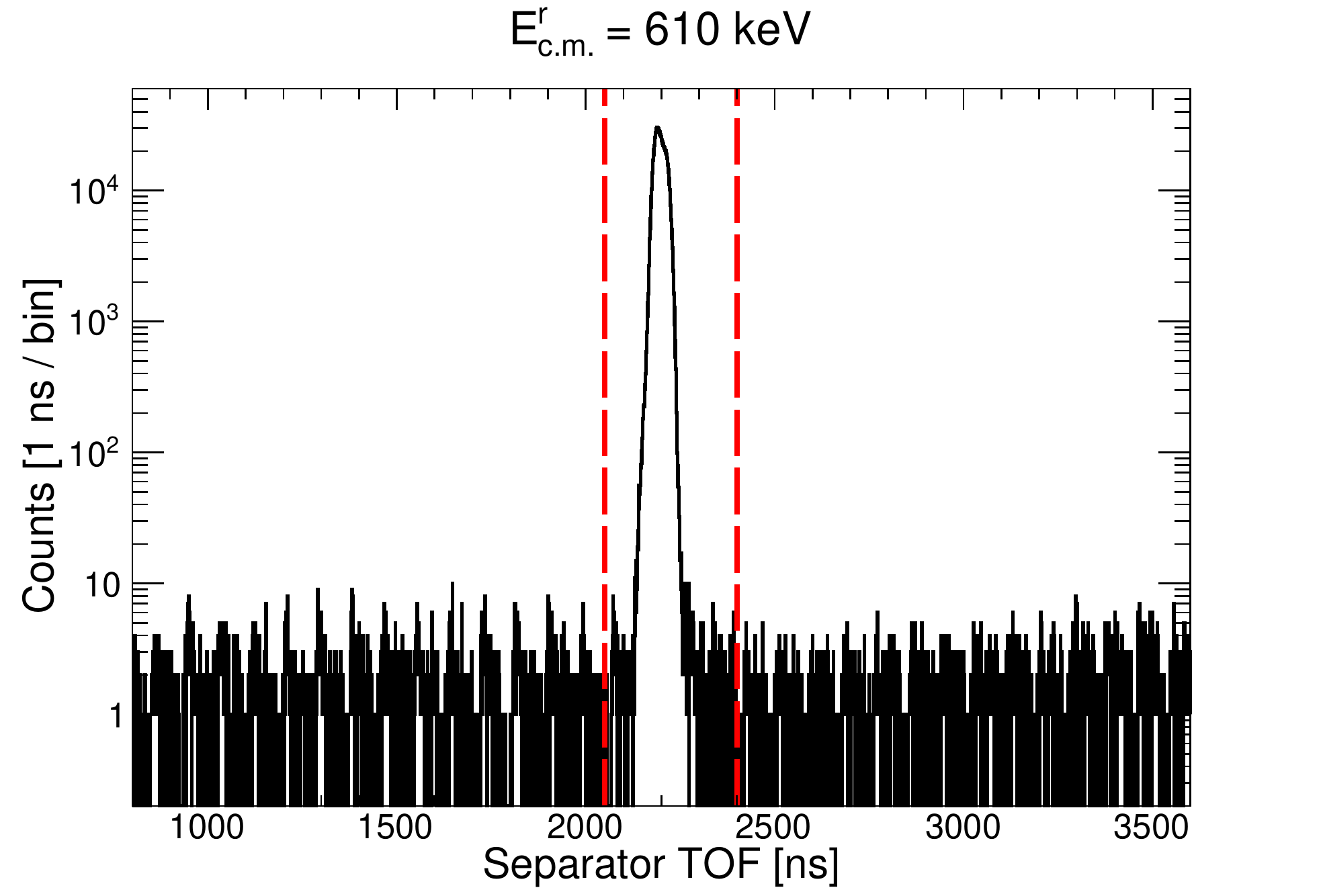}
\endminipage \\
\minipage{0.33\textwidth}
  \includegraphics[width=1.1\textwidth]{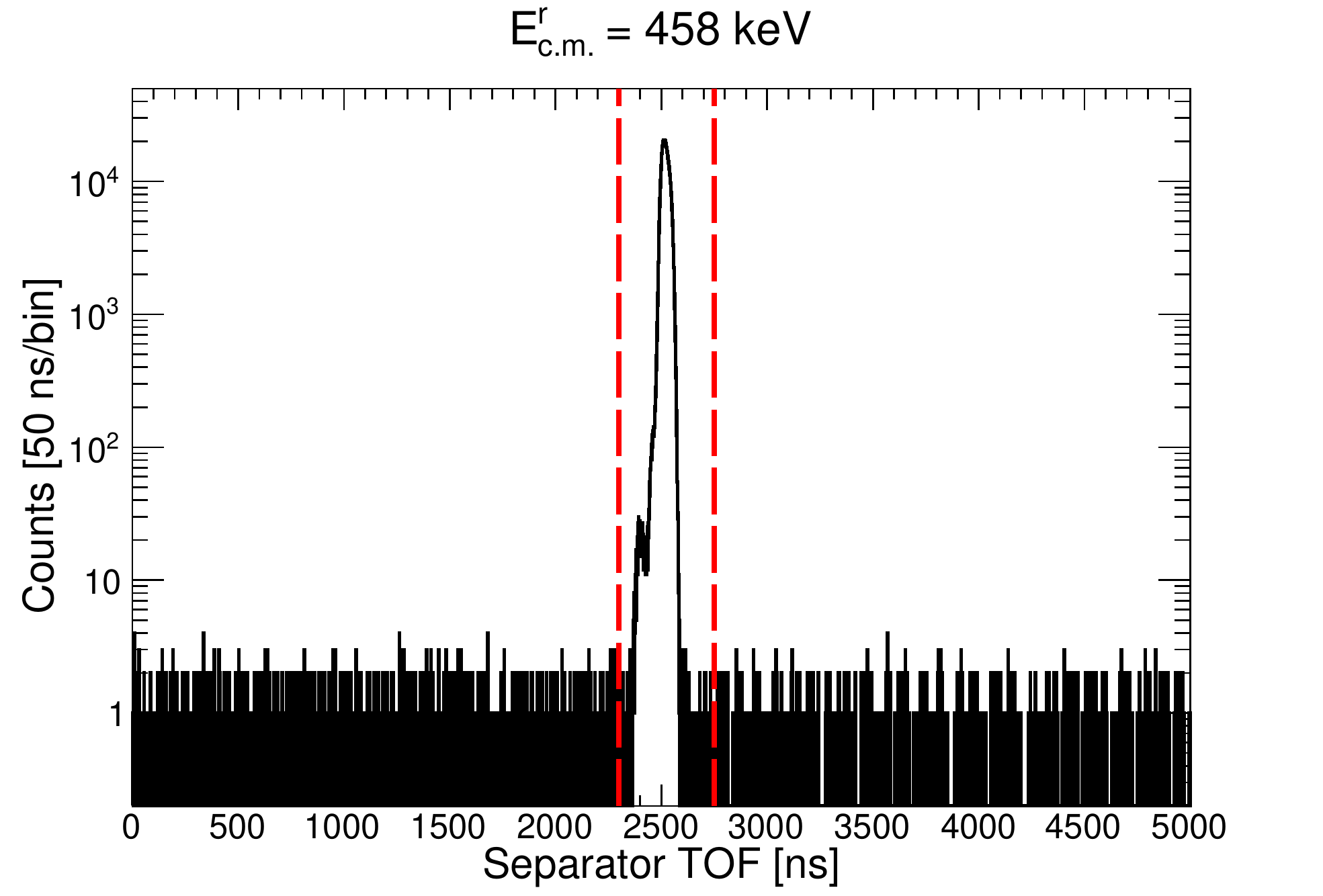}
\endminipage\hfill
\minipage{0.33\textwidth}
  \includegraphics[width=1.1\textwidth]{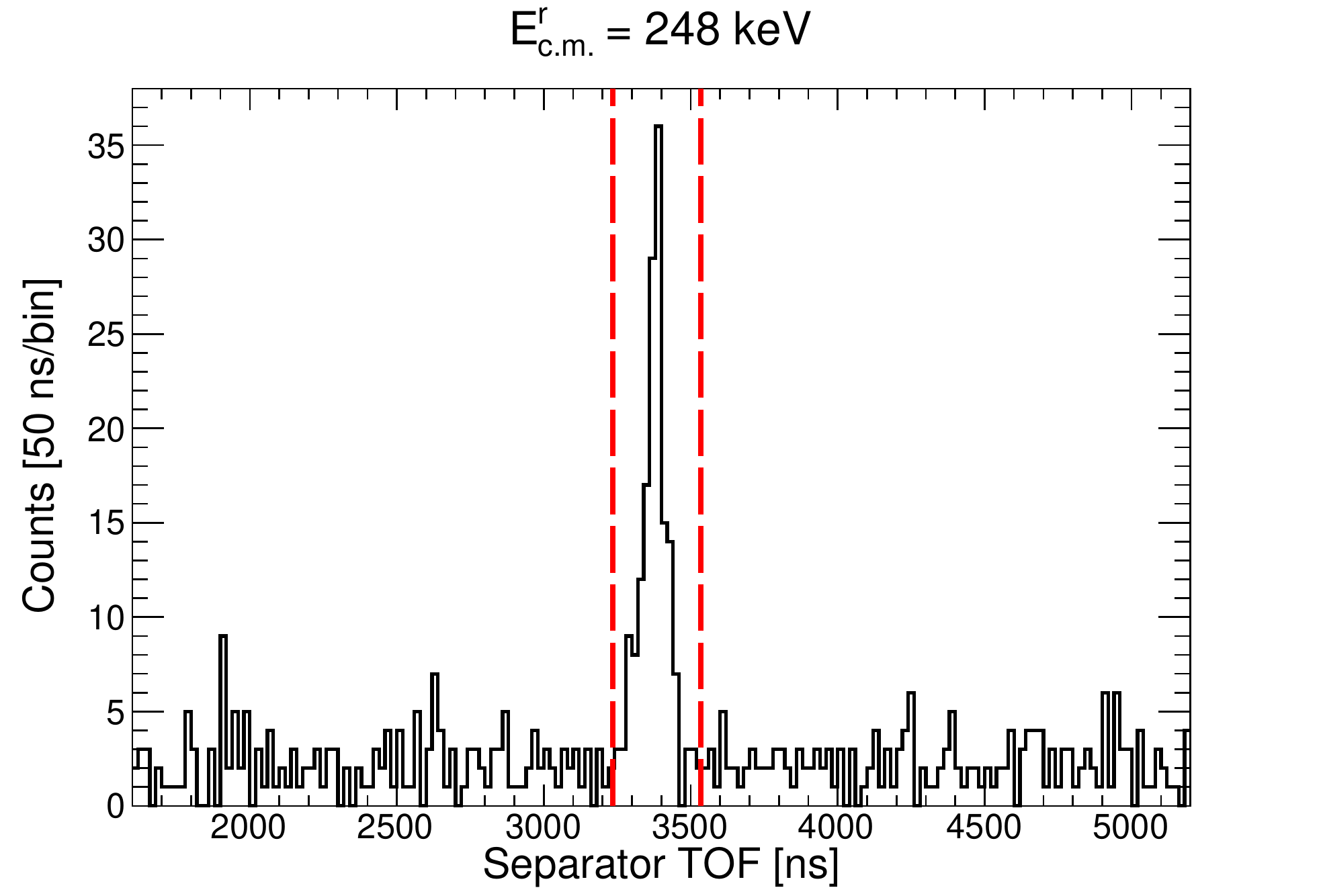}
\endminipage\hfill
\minipage{0.33\textwidth}
  \includegraphics[width=1.1\textwidth]{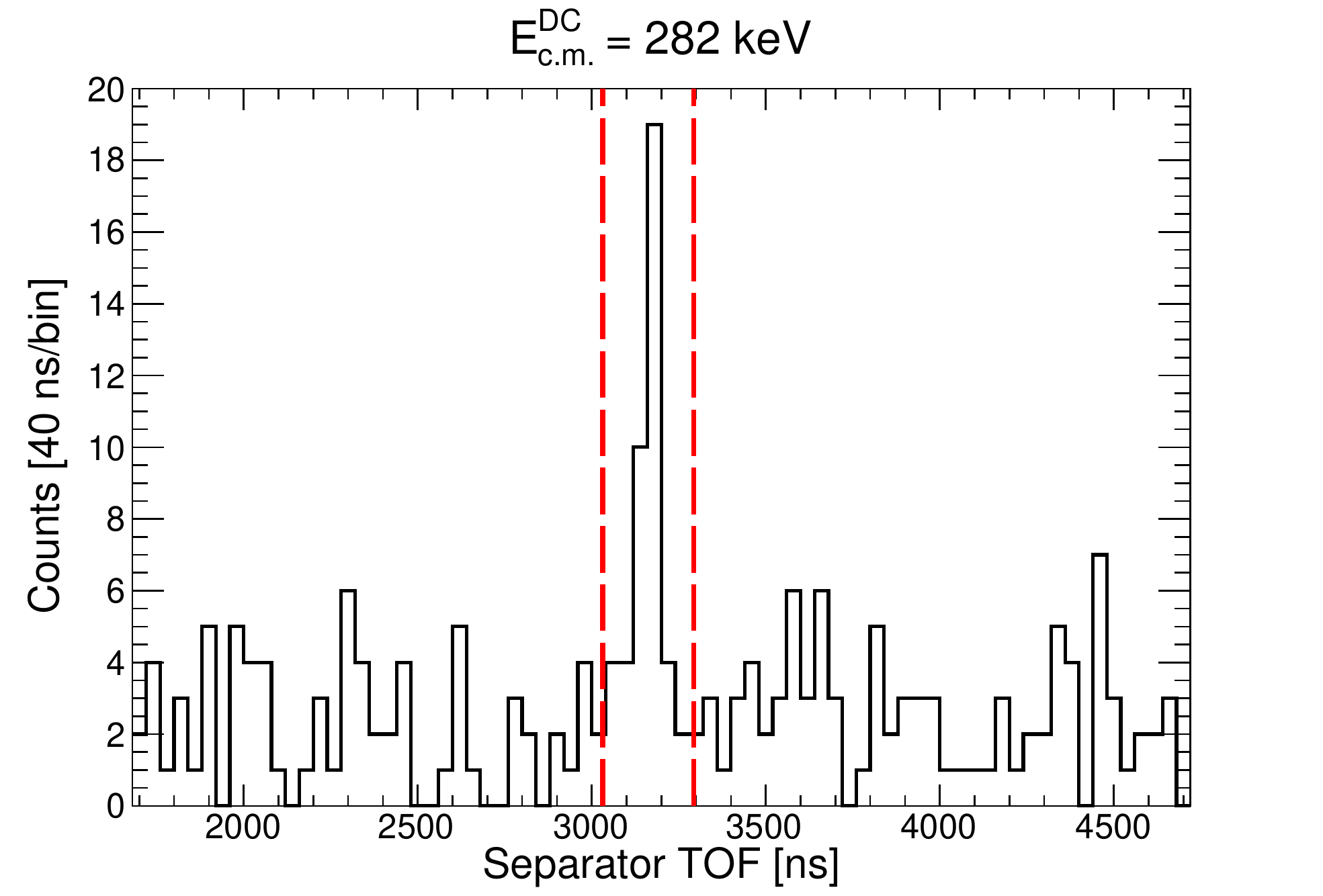}
\endminipage
\caption{Separator time-of-flight (TOF) spectra for resonant yield measurements at $E^{r}_{c.m.} =$ 1222 keV (top-left), 632 keV (top-centre), 610 keV (top-right), 458 keV (bottom-left), and 248 keV (bottom-centre). The spectrum shown in the bottom-right panel pertains to the lowest energy non-resonant yield measurement at $E_{c.m.} = 282$ keV. The separator TOF is constructed from the time difference between a `head'-event recorded by the BGO array and a `tail'-event recorded by any of the focal plane detectors.The background rate within the signal region, bound by the vertical red dashed lines, was estimated by sampling the uniform background outside of the signal region.}
\label{fig:septof}
\end{figure*}


\subsection{\label{sec:res1222}Resonance at $E_{c.m.}=1222$ keV}

The first absolute \Nepg~resonance strength measurement was reported by Keinonen \textit{et al.} \cite{Keinonen-PRC15-1977} for the 1222-keV resonance, with a quoted strength value of $\omega\gamma_{1222} = 10.5 \pm 1.0$ eV. More recently, a study performed at Helmholtz-Zentrum Dresden-Rossendorf measured the ratio of the 1222-keV to 458-keV resonance strengths \cite{Depalo-PRC92-2015}. In that work, the strength of the 1222-keV resonance was reported as $\omega\gamma_{1222} = 11.03 \pm 1.00$ eV, assuming a target thickness derived from a 458-keV resonance strength of $\omega\gamma_{458} = 0.605 \pm 0.062$ eV Here we report a new absolute yield measurement for the 1222-keV resonance that is determined independently of other resonance strength values. 

\begin{figure*}[ttt]
\minipage{0.33\textwidth}
  \includegraphics[width=1.1\textwidth]{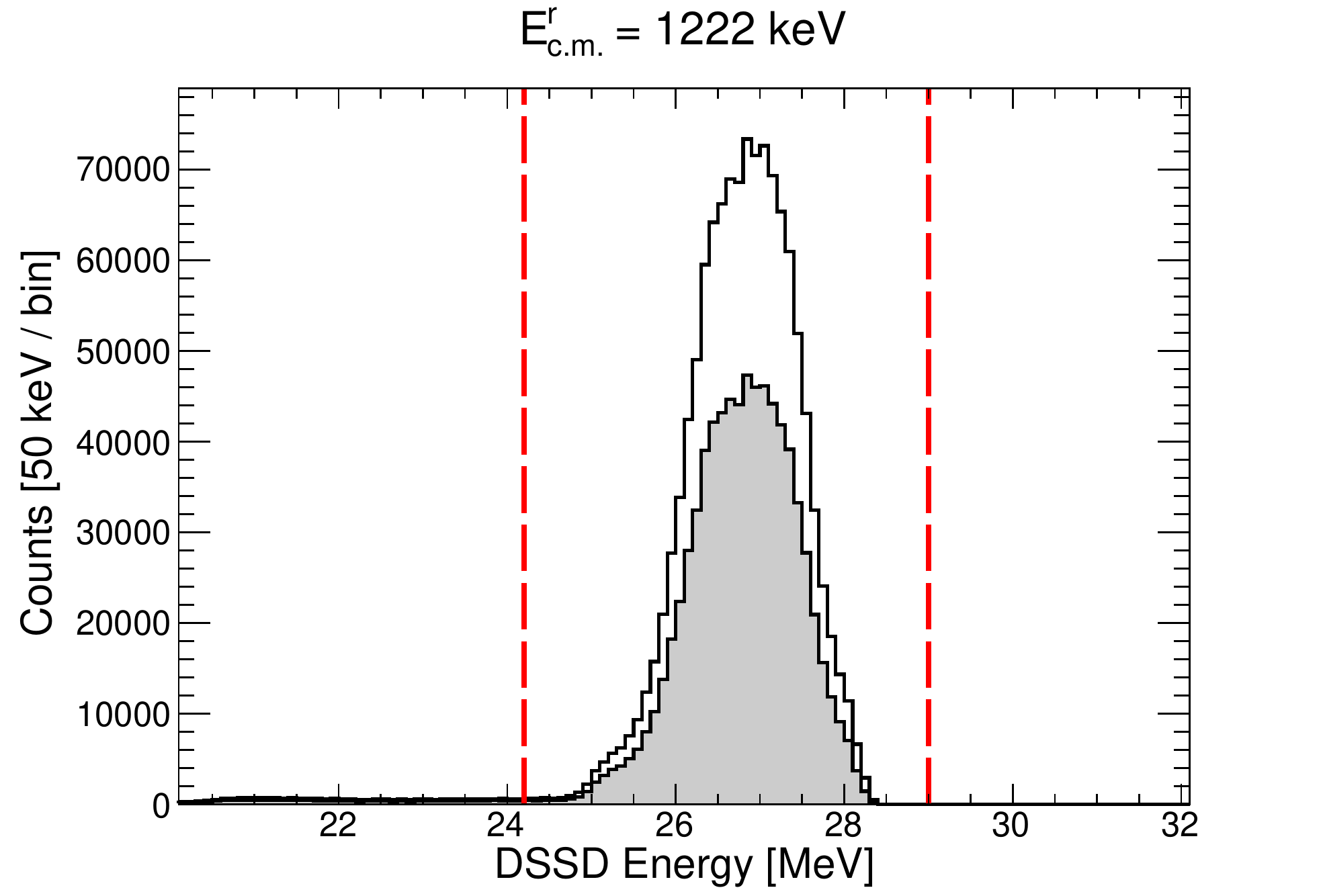}
\endminipage\hfill
\minipage{0.33\textwidth}
  \includegraphics[width=1.1\textwidth]{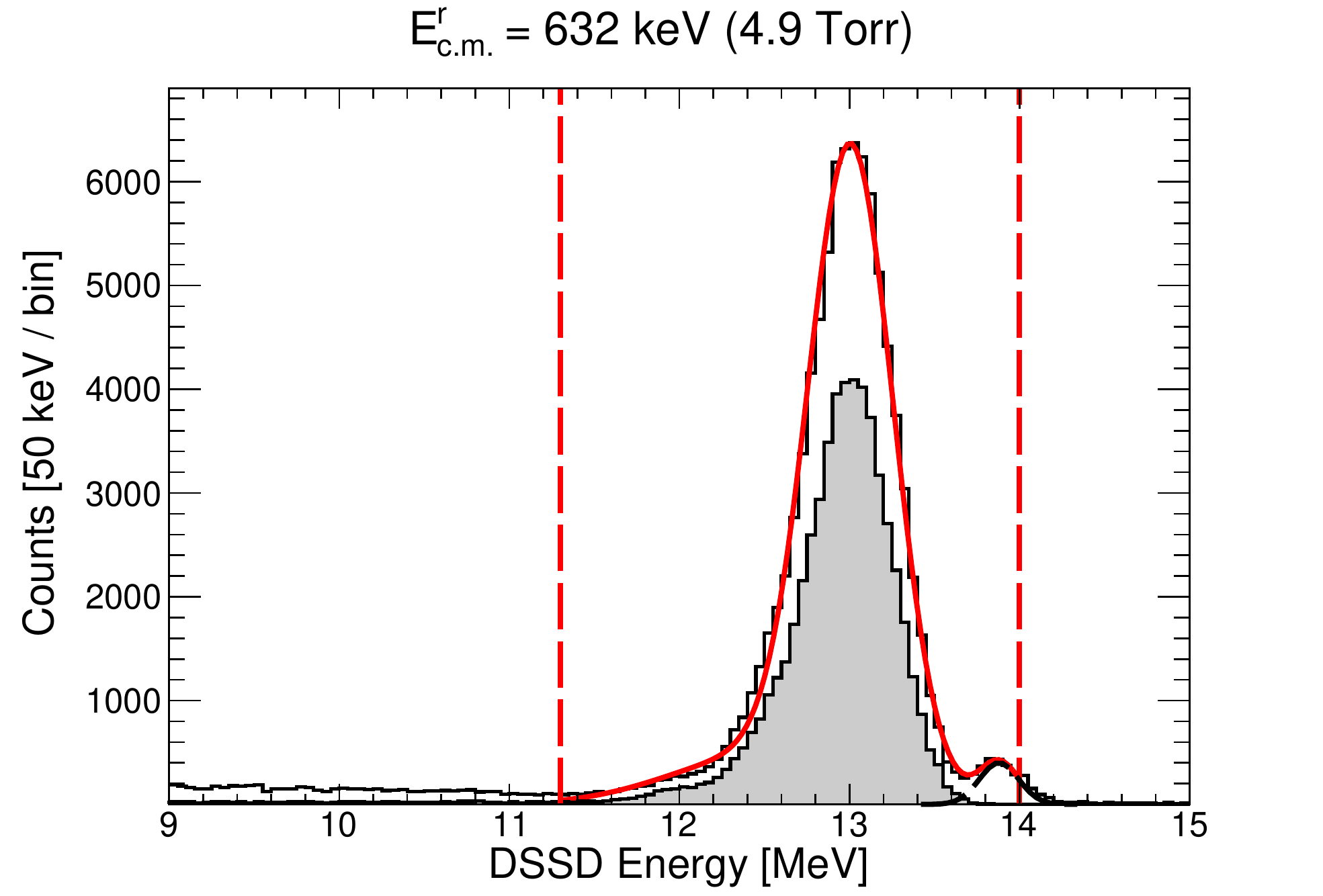}
\endminipage\hfill
\minipage{0.33\textwidth}
  \includegraphics[width=1.1\textwidth]{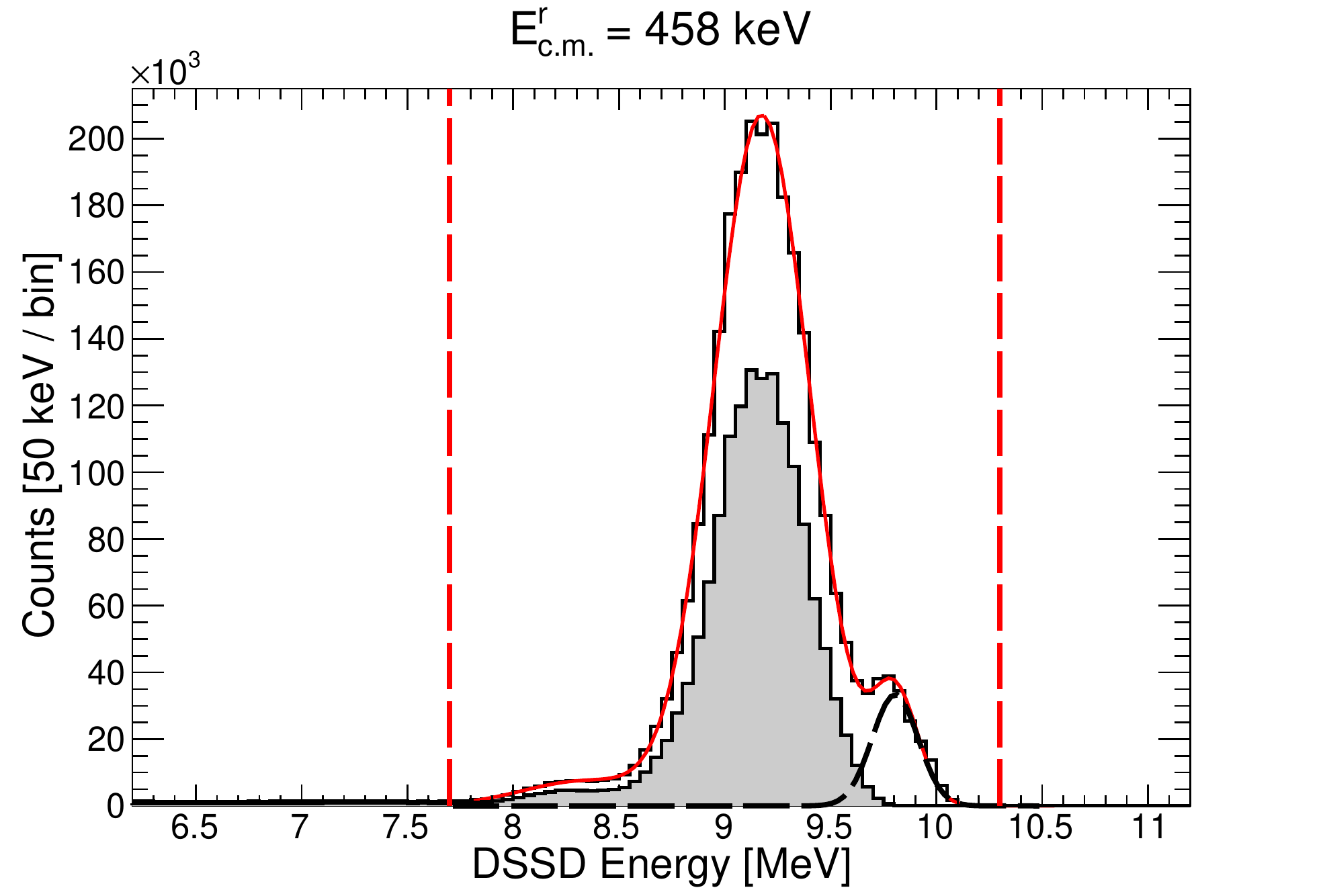}
\endminipage
\caption{DSSD front strip energy spectra for the yield measurements at E$_{{\text c.m.}} = $ 1222~keV (left), 632~keV (centre) and 458~keV (right). Both singles (back line histograms) and coincidence (gray filled histograms) events are shown. The vertical red dashed lines indicate the DSSD energy cuts imposed on the data. The spectra for the 632 keV and 458 keV yield measurements exhibit some leaky-beam contamination at the focal plane; as evidenced by the small peak at higher energy compared to the more prominent recoil peak. These events are entirely suppressed in coincidences, as one would expect. In order to calculate the singles yield, these events were subtracted from the total number of recoils by fitting to a Gaussian (black dashed curve) and integrating over the cut region. The energy cut is slightly expanded to lower energies to include recoils that exhibit additional energy loss through the aluminium contact grid covering the DSSD surface \cite{tengblad2004}. The red solid lines on the $E_{cm} = 632$ and 458 keV plots represents a triple-Gaussian fit across the signal region, accounting for all the aforementioned features in the singles data. Such a fit was unnecessary to perform on the 1222 keV data as leaky-beam contamination was negligible, meaning that no background subtraction was required.}
\label{fig:dssd}
\end{figure*}

 Beam suppression was optimal for this yield measurement, meaning that $^{23}$Na recoils could be easily identified using only the focal plane DSSD, without the need for an accompanying $\gamma$-ray detected in coincidence. Nonetheless, it is useful to first gate on the characteristic separator time-of-flight signal, i.e. the time between a $\gamma$-ray and heavy-ion event, in order to identify the region of interest in the DSSD. The separator TOF gate for this resonance is shown in the top-left panel of Figure \ref{fig:septof}. The DSSD energy spectrum, obtained from both singles only events, and coincidence events gated in the separator TOF signal, is displayed on the left panel of Figure \ref{fig:dssd}. The DSSD spectra appears free from any leaky-beam contamination for both singles and coincidence events. The small tail on the low energy side of the peak is attributed to additional energy loss of recoils traversing the grid of aluminium contacts on the DSSD \cite{tengblad2004}. The imposed cut includes these events, and so the DSSD geometric efficiency of $96.15 \pm 0.5 \%$ is used to account for inter-strip events \cite{Wrede-NIMB204-2003}.
 
 From singles data, we extract a resonance strength of $12.7 \pm 0.7$ eV. In coincidences, assuming primary $\gamma$-ray branching ratios listed on the NNDC database \cite{NNDC}, we recover a resonance strength of $11.7 \pm 1.4$ eV, in good agreement with the singles data. We therefore calculate a weighted average between the present and literature values to give an adopted strength value of $\omega\gamma_{1222} = 11.7 \pm 0.5$ eV.
 
 The 1222-keV resonance strength has a strong impact on the high temperature behaviour of the \Nepg $\;$ reaction rate, with many resonances above 600 keV normalised to this resonance. In calculating the present rate all the resonances in the STARLIB-2013 compilation which are noted as being measured relative to the 1222-keV resonance have been re-normalized to the adopted value.
 
\subsection{\label{sec:res632}Resonance at $E_{c.m.}=632$ keV}

This resonance was initially measured by Meyer \textit{et al.} \cite{Meyer-NPA205-1973} relative to the 610-keV resonance, from which a strength value of $0.285 \pm 0.086$ eV is obtained, in reference to an assumed 610-keV resonance strength of $2.2 \pm 0.5$ eV. This is significantly larger than results from a more recent study by Depalo \textit{et al.}, who report a resonance strength of $0.032^{+0.024}_{-0.009}$ eV for the 632-keV resonance \cite{Depalo-PRC92-2015}. This was also a relative measurement, utilizing the 1222-keV and 458-keV resonances as references. The authors speculate that the original study by Meyer \textit{et al.} may have been affected by contamination from the strong neighbouring resonance at 610 keV. To be sure that the present work is free from such contamination we performed separate yield measurements for this resonance at three different gas target pressures. If there were multiple resonances present in the target then one would expect to find some pressure dependence on the calculated yield and measured resonance energy.

\begin{table}[h!]
\caption{Resonance energies and strengths derived from singles and coincidence data for the $E_{c.m.} = 632$ keV resonance at three different gas pressures. No dependency is observed on the resonance strengths or energies with respect to target pressure. The resonance energies were determined from the BGO hit pattern method described in Ref \cite{Hutcheon-NIMPRA-2012}. Note that the $10\%$ systematic uncertainty related to simulated BGO efficiency has been factored out of the coincidence resonance strengths to allow better point-to-point comparison at the different target pressures.} \smallskip
\centering
\begin{tabular}{l c c c}
	\toprule[1pt]\midrule[0.3pt]
	\noalign{\smallskip}
	Pressure (Torr) & $E_{c.m.}^{r}$ (keV)  &\multicolumn{2}{c}{$\omega\gamma$ (eV)}
	\\  \cline{3-4}
	\noalign{\smallskip}
	 & & Singles & Coincidences  \\ \cline{1-4}
	\noalign{\smallskip}
	4.871(3) & $631.7 \pm 0.1$ & $0.476 \pm 0.033$ & $0.477 \pm 0.034$ \\
	3.169(3) & $632.1 \pm 0.1$ & $0.454 \pm 0.027$ & $0.422 \pm 0.027$ \\
	2.205(8) & $632.0 \pm 0.1$ & $0.496 \pm 0.034$ & $0.468 \pm 0.033$ \\
	\noalign{\smallskip} 
	\midrule[0.3pt]\bottomrule[1pt]
\end{tabular}
\label{table:results_632res}
\end{table}

Table \ref{table:results_632res} lists the calculated strengths for the three different target pressures; no pressure dependence on the yield is evident. This would not be the case if a contaminating resonance were on the periphery of the gas target energy coverage. Moreover, from the measured energy loss across the target of 14.75 keV (center of mass) at the highest gas pressure used, the 610-keV resonance would be located some 12.2 cm downstream of the end of the gas target. In addition to the yield, one would also expect the calculated resonance energy to be affected by pressure changes. The resonance energies, determined via the BGO hit pattern method detailed in Ref \cite{Hutcheon-NIMPRA-2012}, are also given in Table \ref{table:results_632res} and all agree on a resonance energy of $631.7(4)$ keV based on an unweighted average over each measurement. Taking all this information together,  we conclude that our resonance strength is not being influenced by contamination, and adopt a final strength value of $\omega\gamma_{632} = 0.472 \pm 0.018$ eV based on a weighted average of the singes resonance strengths listed in Table \ref{table:results_632res}.

The discrepancy between the present work with respect to the result by Depalo \textit{et al.} \cite{Depalo-PRC92-2015} is not easily reconciled, since many systematic effects would produce similar discrepancies for other resonances where reasonable agreement is found. With regards to the previous Meyer \textit{et al.} value \cite{Meyer-NPA205-1973}, the disagreement here is lessened slightly by re-normalising to the 610-keV resonance strength proposed in this work, which results in a $\approx11\%$ increase in the strength to $0.324 \pm 0.099$ eV, bringing it to within $2\sigma$ agreement with respect to the present value. However, taking the literature into consideration as a whole we do not recommend this as a reference resonance.

\subsection{\label{sec:res612}Resonance at $E_{c.m.}=610$ keV}

The narrow resonance at $E_{c.m.}=610$ keV was measured relative to the 1222-keV resonance by Keininen \textit{et al.} \cite{Keinonen-PRC15-1977}. In that study, a resonance strength value of $\omega\gamma_{610} = 2.8 \pm 0.3$ eV was reported for the $E_{c.m.} = 610$ keV resonance. This is in agreement with an earlier reported measurement by Meyer \textit{et al.} \cite{Meyer-NPA205-1973}. More recently, this resonance was amongst those targeted by Depalo \textit{et al.} \cite{Depalo-PRC92-2015}. Again, the reported strength in that study is measured relative to both the 458-keV and 1222-keV resonances. The adopted value is calculated from the weighted average of the two relative measurements, quoted as $\omega\gamma_{610} = 2.45 \pm 0.18$ eV, which is good agreement with the preexisting literature. 

Unfortunately, excessive leaky-beam background prevented a result from being extracted using singles data. Instead, we obtain a coincidence result of $\omega\gamma_{610} = 2.44 \pm 0.32$ eV. For obtaining the BGO coincidence efficiency we utilized the branching ratios published by Depalo \textit{et al.} as inputs to the \textsc{GEANT3} simulation. Our result is well within $1\sigma$ agreement of all values available in the literature. Therefore, we propose to adopt a weighted average of all the literature values as $\omega\gamma_{610} = 2.50 \pm 0.13$ eV. Given the good agreement in the literature on the strength of this resonance, we propose the $E^{r}_{c.m.} = 610$ keV resonance as a good reference resonance. 

\subsection{\label{sec:res458}Resonance at $E_{c.m.}=458$ keV}

The general lack of well-measured reference resonances was commented upon by Longland \textit{et al.} \cite{Longland-PRC81-2010}, particularly for noble gas targets where issues related to target stochiometry can be especially troublesome. In the particular case of \Nepg~ accurate reference resonances will also be of interest for other reaction rate studies. For instance, studies of $^{22}$Ne$~+~\alpha$ reaction rates, which are of importance for the weak s-process, have used \Nepg~ resonances to normalize target thickness \cite{Giesen-NPA561-1993}. The isolated narrow \Nepg \; resonance at $E^{r}_{c.m.} =$ 458 keV is a potentially advantageous candidate to use as a reference due to its relatively large strength and location at a moderately accessible energy.

\begin{figure}[h!]
 \includegraphics[width=0.45\textwidth]{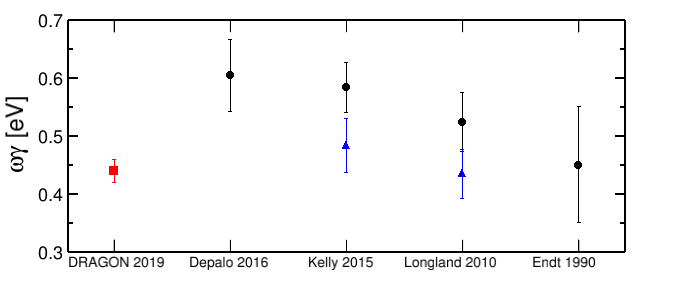}
\caption{Comparison between present and literature values for the 458 keV resonance. The blue solid triangles indicate re-normalized values for the Kelly \cite{Kelly-PRC92-2015} and Longland \cite{Longland-PRC81-2010} measurements if one instead adopts the $E_{r} = 394$ keV $^{27}$Al$+p$ reference resonance reported in Ref \cite{Harissopulos-EPJA9-2000}, as opposed to the strength reported in Ref \cite{Powell-NPA644-1998} (see text for details).} \label{fig:compare458}
 \end{figure}

This resonance was measured over the course of $\sim6.5$ hours of data-taking, with a total estimated number of $(7.991 \pm 0.092) \times 10^{15}$ $^{22}$Ne beam ions incident on target. A total of $(2.1923 \pm 0.0018) \times 10^{6}$ singles and $(1.2780 \pm 0.0011) \times 10^{6}$  coincident recoils were recorded (for a BGO threshold of $E_{\gamma}^{(0)} \geqslant 2$ MeV in the case of coincidences). The ability to accept such high intensity beams with excellent background suppression is a key advantage of the DRAGON facility; allowing high statistics results with little required measurement time. Figure \ref{fig:compare458} shows a comparison between the present work and the literature. Here we find significant disagreement with respect to recent measurements by Depalo \textit{et al.} \cite{Depalo-PRC92-2015} and Kelly \textit{et al.} \cite{Kelly-PRC92-2015}, both of which are higher than the present value of $\omega\gamma_{458} = 0.44 \pm 0.02$ eV, which we calculate based on a weighted average of singles and coincidence measurements. For the $\gamma$-recoil coincidence efficiency we utilized the recommended branching ratios by Kelly \textit{et al.} \cite{Kelly-PRC92-2015} as inputs for the \textsc{GEANT3} simulation.

\begin{table*}
\centering
\caption{Table of \Nepg~resonances used for calculating the thermonuclear reaction rate. Literature values are also listed for comparison. Adopted strength values are calculated from weighted averages of the literature and both single and coincidence results from the present work, except for the resonances at $E_{c.m.} = 458$ and 632 in which we find significant disagreement with respect to the literature. For these we adopt a weighted average of the singles and coincidence results. Resonances located between $E_{c.m.} = 632$ and 1222 keV, and beyond 1222 keV, are adopted from Ref \cite{Iliadis-NPA841-2010} unchanged and so are not included in the table.} \smallskip
\begin{tabular}{l c c c c c}
	\toprule[1pt]\midrule[0.3pt]
	\noalign{\smallskip}
	 $E^{r}_{c.m.}$ (keV) & \multicolumn{4}{c}{$\omega\gamma$ (eV)} & Screening \\ \noalign{\smallskip} \cmidrule[0.3pt]{2-5} \noalign{\smallskip}
	 & Literature & \multicolumn{2}{c}{Present Work} & Adopted & enhancement \\ \noalign{\smallskip} \cmidrule[0.3pt]{3-4} \noalign{\smallskip}
	&  & Singles & Coincidences & & factor, $f$ \\ \noalign{\smallskip} \hline \noalign{\smallskip}
	35 & $(3.1 \pm 1.2) \times 10^{-15}$ &  &  & $(3.1 \pm 1.2) \times 10^{-15}$ & \\
	68 & $\leqslant 1.5 \times 10^{-9}$ \cite{Cavanna-PRL115-2015,Depalo-PRC94-2016} &  &  &  &  \\
	& $\leqslant 6 \times 10^{-11}$ \cite{Ferraro-PRL-2018} & & & & \\
	100 & $\leqslant 7.6 \times 10^{-9}$ \cite{Cavanna-PRL115-2015,Depalo-PRC94-2016} & &  &  & \\
	& $\leqslant 7 \times 10^{-11}$ \cite{Ferraro-PRL-2018} & & & & \\
	$149$ & $(1.8 \pm 0.2) \times 10^{-7}$ \cite{CavannaError} &  & $1.7~^{+0.5}_{-0.4} \times 10^{-7}$ &  $(1.9 \pm 0.1) \times 10^{-7}$ & 1.074 \\
	& $(2.0 \pm 0.4) \times 10^{-7}$ \cite{Kelly-PRC95-2017} & & & & \\
	& $(2.2 \pm 0.2) \times 10^{-7}$ \cite{Ferraro-PRL-2018} & & & & \\
	$181$ & $(2.2 \pm 0.2) \times 10^{-6}$ \cite{CavannaError} &  & $(2.2 \pm 0.4) \times 10^{-6}$ &  $(2.3 \pm 0.1) \times 10^{-6}$ & 1.055 \\
	& $(2.3 \pm 0.3) \times 10^{-6}$ \cite{Kelly-PRC95-2017} & & & & \\
	& $(2.7 \pm 0.2) \times 10^{-6}$ \cite{Ferraro-PRL-2018} & & & & \\
	$248$ & $(8.2 \pm 0.7) \times 10^{-6}$ \cite{CavannaError} & & $(8.5 \pm 1.4) \times 10^{-6}$ &  $(8.9 \pm 0.5) \times 10^{-6}$ & 1.034 \\
	& $(9.7 \pm 0.7) \times 10^{-6}$ \cite{Ferraro-PRL-2018} & & & & \\
	417 & $(7.9 \pm 0.6) \times 10^{-2}$ \cite{Depalo-PRC92-2015} &  &  & $(8.2 \pm 0.5) \times 10^{-2}$ & \\
	& $(8.8 \pm 1.0) \times 10^{-2}$ \cite{Kelly-PRC95-2017} & & & & \\
	458 & $(5.8 \pm 0.4) \times 10^{-1}$ \cite{Kelly-PRC92-2015} & $(4.4 \pm 0.2) \times 10^{-1}$ & $(4.4 \pm 0.5) \times 10^{-1}$ & $(4.4 \pm 0.2) \times 10^{-1}$ & \\
	& $(6.1 \pm 0.6) \times 10^{-1}$ \cite{Depalo-PRC92-2015} & & & & \\
	610 & $2.8 \pm 0.3$ \cite{Keinonen-PRC15-1977} & & $2.44 \pm 0.32$ & $2.50 \pm 0.13$ & \\
	& $2.45 \pm 0.18$ \cite{Depalo-PRC92-2015} & & & & \\
	632 & $(2.85 \pm 0.86) \times 10^{-1}$ \cite{Meyer-NPA205-1973} & $(4.7 \pm 0.2) \times 10^{-1}$ & $(4.5 \pm 0.3) \times 10^{-1}$ & $(4.7 \pm 0.2) \times 10^{-1}$ & \\
	& $3.2 ^{+2.4}_{-0.9} \times 10^{-2}$ \cite{Depalo-PRC92-2015} & & & & \\
	1222 & $10.5 \pm 1.0$ \cite{Keinonen-PRC15-1977} & $12.7 \pm 0.7$ & $11.7 \pm 1.4$ & $11.7 \pm 0.5$ & \\
	& $11.0 \pm 1.0$ \cite{Depalo-PRC92-2015} & & & & \\
	\noalign{\smallskip} 
	\midrule[0.3pt]\bottomrule[1pt] 
\end{tabular}
\label{table:results}
\end{table*}

Given the discrepancy between the present work and the literature, it is necessary to revisit the techniques employed to derive these strength values. The result presented by Kelly \textit{et al.} \cite{Kelly-PRC92-2015} is obtained by applying an updated direct-to-ground state branching ratio to the previous measurement by Longland \textit{et al} \cite{Longland-PRC81-2010}, which used this branch to obtain the \Nepg \; yield. Longland \textit{et al.} determined the strength of the 458-keV resonance via a novel technique involving depth profiling the neon target content implanted into an aluminium substrate. Utilizing the $E_{r} = 394$-keV $^{27}$Al$(p,\gamma)^{28}$Si resonance strength reported in Ref \cite{Powell-NPA644-1998}, a profile of the target stoichiometry was obtained by fitting the $^{27}$Al$(p,\gamma)^{28}$Si resonance yield. However, we note that the $E_{r} = 394$-keV $^{27}$Al + p resonance, from which the yield was used to determine the target stoichiometry, has a lower strength than that reported in a more recent measurement by Harissopulos \textit{et al.} \cite{Harissopulos-EPJA9-2000}. If one assumes that the depth-profiling techniques allows for a simple re-normalisation of the target content then, after applying the new direct-to-ground state branch from Kelly \textit{et al.} \cite{Kelly-PRC92-2015}, the strength from that work becomes $\omega\gamma_{458} = 0.484 \pm 0.052$ eV, in agreement with the present value. This inter dependency of relative strength measurements emphasizes the case for absolute techniques to precisely measure candidate reference resonances, a task for which DRAGON is well suited. 



\subsection{\label{sec:res249}Resonance at $E_{c.m.}=248$ keV}

This resonance was amongst the three low energy resonances reported by the LUNA collaboration \cite{Cavanna-PRL115-2015,Depalo-PRC94-2016}. Here we report a strength value that lies between the first LUNA measurement and a more recent study by the LUNA group \cite{Ferraro-PRL-2018}, thus supporting a larger strength than previous upper limits \cite{Gorres1982,Hale2002}.

The lower-centre panel of Figure \ref{fig:septof} shows the separator TOF spectrum for this yield measurement. The small background under the indicated signal region is due to random coincidences between background $\gamma$-rays and scattered leaky-beam making it to the focal plane. The background contribution was evaluated by sampling counts in 10 equal sized regions above and below the signal region to obtain a mean background expectation value. This was then subtracted from the signal to give the final number of recoils, with $1\sigma$ confidence bounds calculated using the Rolke method \cite{rolke2001} assuming a Poisson background model. From this we find a resonance strength of $\omega\gamma_{248} = 8.5 \pm 1.4~\mu$eV. The coincidence efficiency was obtained through simulation, using primary branching ratios published by Depalo \textit{et al.} \cite{Depalo-PRC94-2016}. Unfortunately, no singles result could be extracted due to overwhelming leaky-beam background at the focal plane.

This resonance has only a minor influence on the \Nepg \; rate, with greater contributions derived from the other two low-energy resonances at 181 and 149 keV, as discussed in Section \ref{sec:rate}.

\subsection{\label{sec:res181}Resonance at $E_{c.m.}=181$ keV}

The first direct measurement for the resonance at 181 keV was reported by the LUNA collaboration \cite{Cavanna-PRL115-2015}, and was later confirmed in a measurement at TUNL \cite{Kelly-PRC95-2017}. The latter study measured the resonant yield relative to the 458-keV resonance strength reported in Ref \cite{Kelly-PRC92-2015}. Both the TUNL study and initial LUNA study are in agreement, finding resonance strengths of $2.3 \pm 0.3$ and $2.2 \pm 0.2$ $\mu$eV, respectively, which both lie just below that of the previous direct upper limit of $\leqslant 2.6~\mu$eV set by Gorres \textit{et al.} \cite{Gorres1982}. More recently the LUNA group re-measured this resonance, instead using a different set-up comprising a BGO $\gamma$-ray spectrometer. This study found a larger strength, compared with the previous LUNA measurement, of $2.7 \pm 0.2$ $\mu$ eV, which the authors attribute to greater sensitivity to weak branches that may have been missed by the previous measurement. Here we report a resonance strength of $\omega\gamma_{181} = 2.2 \pm 0.4 ~\mu$eV that is consistent to within $1\sigma$ of all the aforementioned literature values, and also in-keeping with the upper limit set by Gorres \textit{et al.} \cite{Gorres1982}.

\begin{figure}
 \includegraphics[width=0.45\textwidth]{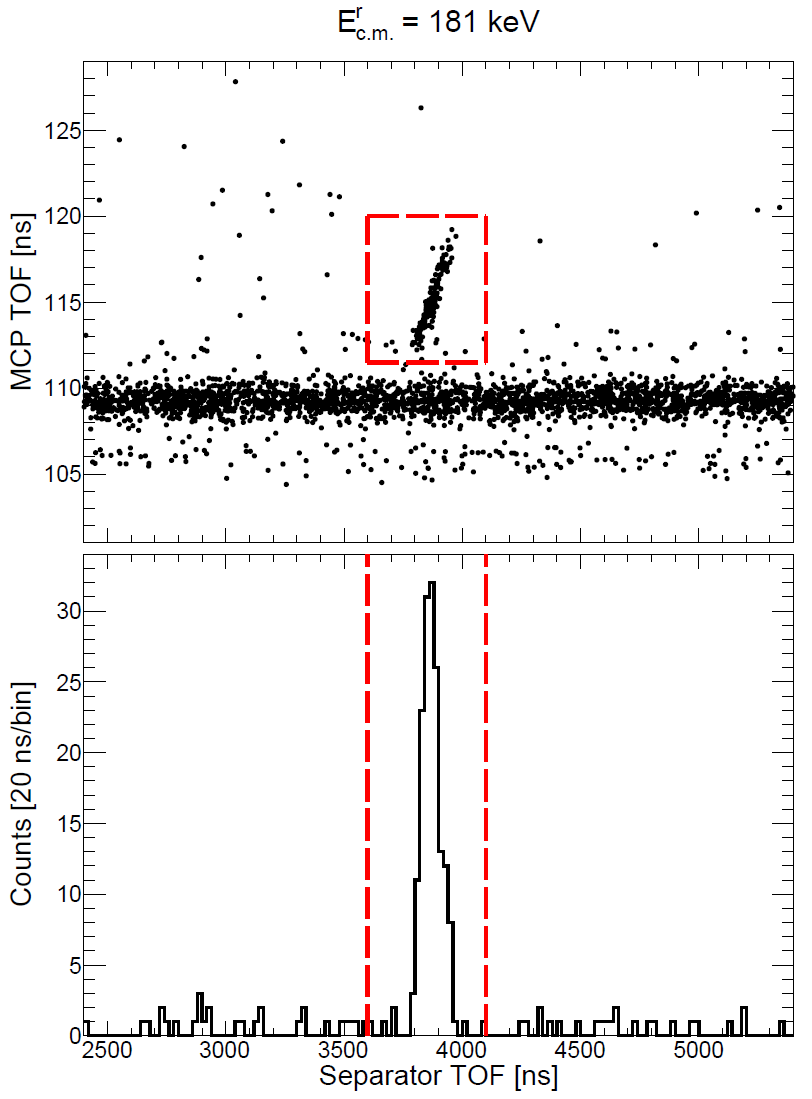}
\caption{(top pannel) MCP TOF vs Separator TOF for the on resonance yield measurement at $E_{c.m.} = 181$ keV, with an applied BGO threshold of $E^{(0)}_{\gamma} \geqslant2$ MeV. The recoil locus is highlighted by the red dashed lines, which constitute the signal timing gates. (bottom pannel) The separator TOF spectrum with applied MCP-TOF gate and $E^{(0)}_{\gamma} \geqslant2$ MeV BGO energy threshold. The background was estimated by taking the average of five sample below the signal region, and five above, each of equal width to the signal gate. The total number of recoils is then given by the total signal, in this case 166 counts, minus a background estimate of 11 counts, which comes to $155~^{+14}_{-12}$ $^{23}$Na recoils collected for this yield measurement} \label{fig:Eres181keV_septof}
 \end{figure}

This yield measurement benefited from the increased selectivity provided by a fully functioning MCP-TOF system, following a recent replacement of both MCP detectors. The MCP vs Separator TOF spectrum is plotted in Figure \ref{fig:Eres181keV_septof}, along with the separator TOF gated on the MCP-TOF signal. A distinct grouping of recoil events is clearly distinguished from leaky-beam background, the latter being uncorrelated with respect to the separator TOF. There is still nonetheless a small background contribution arising within the displayed signal gate that ought to be accounted for. The separator TOF spectrum, gated on the MCP-TOF signal region, is shown in the lower panel of Figure \ref{fig:Eres181keV_septof}. An estimate of the background within the signal region was calculated by sampling the background above and below the separator TOF signal region. This estimate was then subtracted from the total signal, and $1\sigma$ confidence bounds calculated using the Rolke method \cite{rolke2001}. For deriving the coincidence efficiency we utilized the branching ratios published from the TUNL study \cite{Kelly-PRC95-2017}, which reports an additional weak $\gamma$-decay branch to the ground state.

\subsection{\label{sec:res149}Resonance at $E_{c.m.}=149$ keV}

The role this resonance plays in the \Nepg~ reaction rate was initially thought to be minimal, based on an indirect upper limit of $\omega\gamma \leqslant 9.2 \times 10^{-9}$ eV obtained from a $(^{3}$He,$d)$ transfer study conducted by Hale \textit{et al.} \cite{Hale2002}. This upper limit is based on a spectroscopic factor assuming an $L=3$ transfer to the $E_{x} = 8944$-keV state. Despite this study not being able to distinguish between an $L=2$ or $L=3$ transfer, an $L=2$ transfer was discounted based on an assumed spin-parity of $J^{\pi} = 7/2 ^{-}$ for the resonance in question. However, a recent $^{12}$C$(^{12}$C,$p\gamma)^{23}$Na study using Gammasphere revealed that this resonance in fact comprises a doublet: one  $J^{\pi} = 7/2 ^{-}$ state, and a second state at $E_{x} = 8944$ keV with a tentatively assigned $3/2^{+}$ spin-parity \cite{Jenkins-PRC87-2013}. The literature surrounding the spin-parity assignment of this state, and interpretation of transfer data in-lieu of new spectroscopic information, is discussed in detail by Kelly \textit{et al.} \cite{Kelly-PRC95-2017}. 

\begin{figure}
 \includegraphics[width=0.45\textwidth]{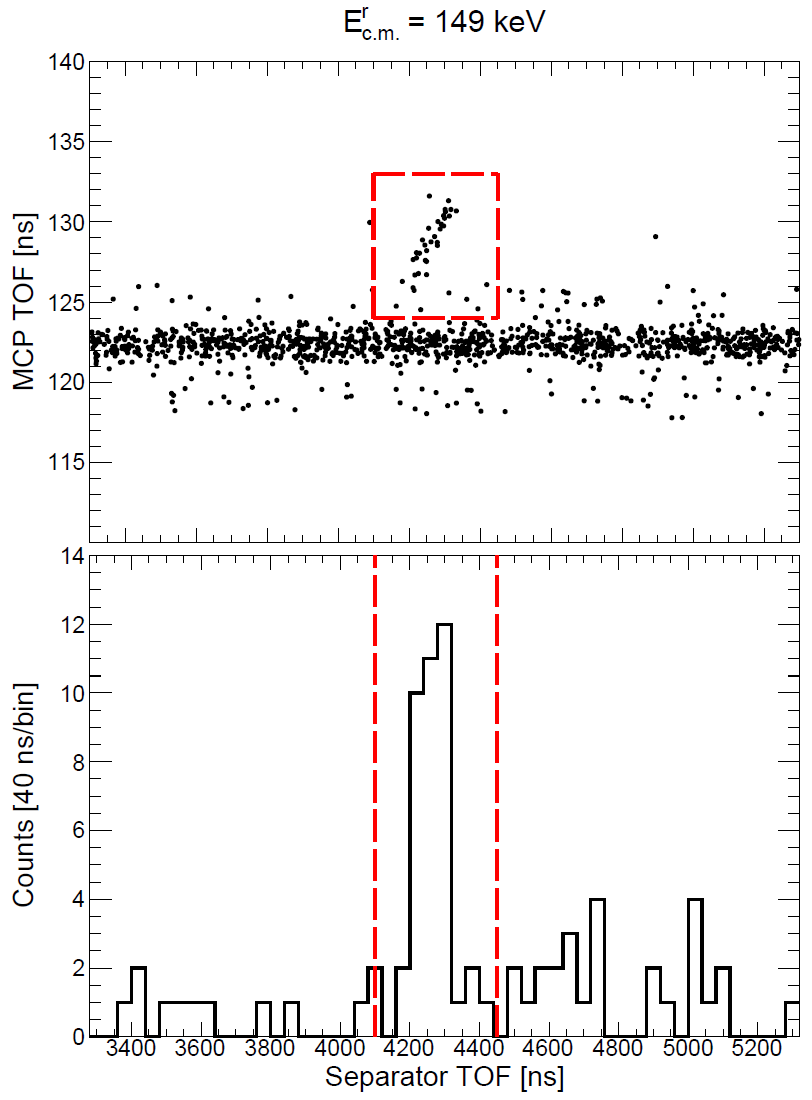}
\caption{(top pannel) MCP TOF vs Separator TOF for the on resonance yield measurement at $E_{c.m.} = 149$ keV, with an applied BGO threshold of $E^{(0)}_{\gamma} \geqslant2.5$ MeV. The recoil locus is highlighted by the red dashed lines, which constitute the signal timing gates. (bottom pannel) The separator TOF spectrum with applied MCP-TOF gate and $E^{(0)}_{\gamma} \geqslant2.5$ MeV BGO energy threshold. The background was estimated by taking the average of five sample below the signal region, and five above, each of equal width to the signal gate. The total number of recoils is then given by the total signal, in this case 39 counts, minus a background estimate of 6 counts, which comes to $33~^{+8}_{-6}$ $^{23}$Na recoils collected for this yield measurement} \label{fig:Eres149keV_septof}
 \end{figure}

\begin{figure}[ht!]
 \includegraphics[width=0.45\textwidth]{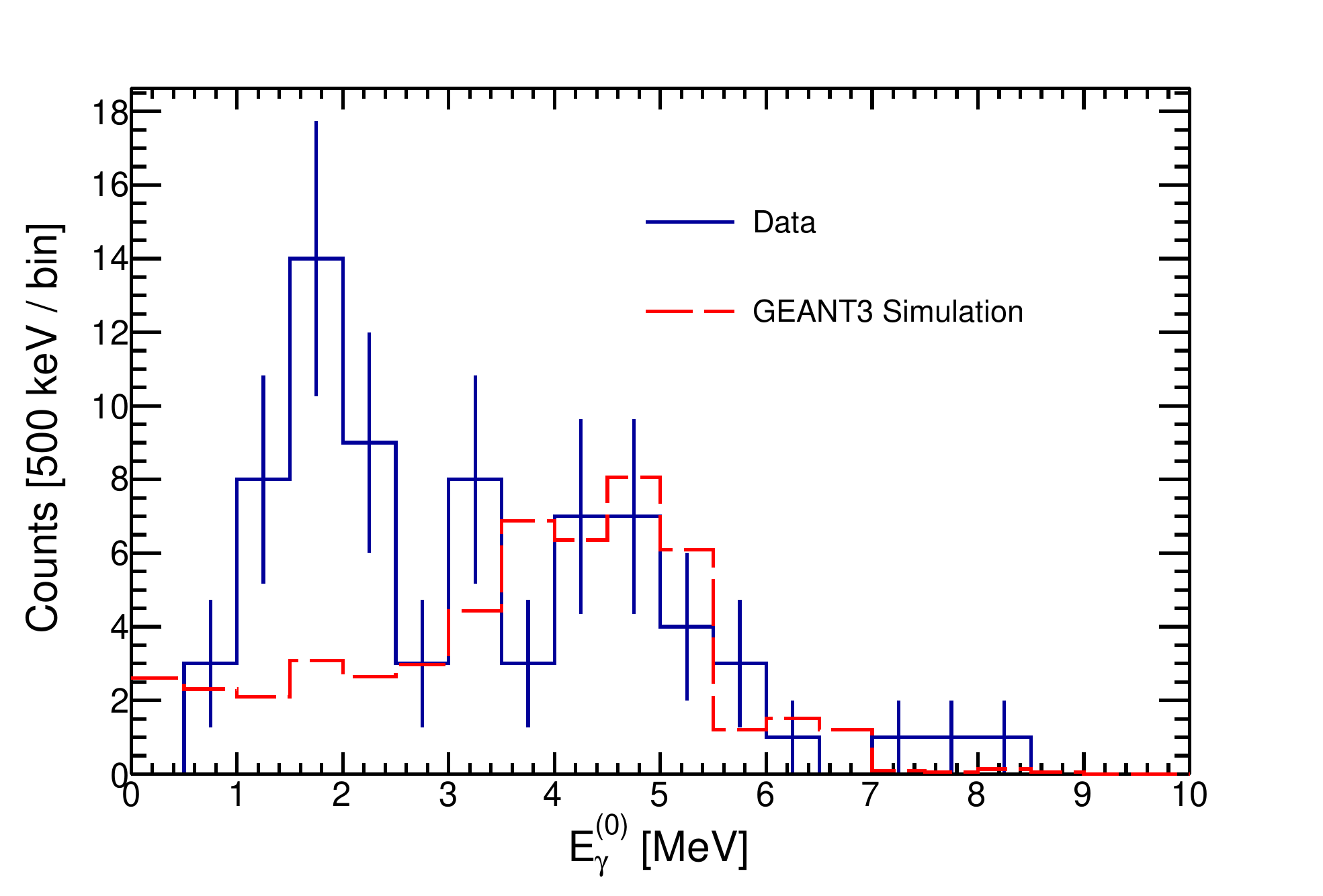}
\caption{Spectrum of highest energy $\gamma$-rays detected by the BGO array $(E^{(0)}_{\gamma})$ in coincidence with a heavy-ion event passing the timing gates for both the MCP-TOF and separator TOF shown in the upper panel in Figure \ref{fig:Eres149keV_septof}. The red dashed line represents simulated data, scaled to the actual data (blue line). The excess of counts at $~2$ MeV, not reproduced in the simulation, are likely being contributed to by random coincidences with background $\gamma$-rays. Therefore, a slightly raised threshold was opted for. Indeed, after performing the separator TOF background subtraction, this threshold choice resulted in a slightly improved statistical error bar compared with a lower 2 MeV threshold used for other yield measurements.} \label{fig:Eres149keV_bgoenergy}
 \end{figure}

In this work we present a new absolute strength measurement for the $E_{c.m.} = 149$ keV resonance. The total number of recoils is obtained in a similar fashion to that explained in the previous section for the $E_{c.m.} = 181$ keV resonance. The separator vs MCP-TOF spectrum is shown on Figure \ref{fig:Eres149keV_septof}, with a $E^{(0)}_{\gamma} \geqslant 2.5$ MeV $\gamma$-ray threshold. The BGO software threshold was optimised during offline analysis by comparing the BGO energy spectrum to simulation; this comparison (shown on Figure \ref{fig:Eres149keV_bgoenergy}) revealed that the signal-to-background could be improved by raising the software-imposed threshold to 2.5 MeV as opposed to a typical 2 MeV threshold used for all other yield measurements. The coincidence efficiency was then obtained for the aforementioned BGO energy gate, assuming the $\gamma$-ray branching ratios put forward by Kelly \textit{et al.} \cite{Kelly-PRC95-2017}. Though within $1\sigma$ agreement, our obtained strength of $\omega\gamma_{149} = 0.17~^{+0.5}_{-0.4}~\mu$eV is lower than those reported by TUNL \cite{Kelly-PRC95-2017} and both LUNA measurements \cite{CavannaError,Ferraro-PRL-2018}. It is perhaps worth noting that for the TUNL result, given that this was measured relative to the 458 keV resonance strength reported in Ref \cite{Kelly-PRC92-2015}, if one were to re-normalise to the 458 keV strength adopted in the present work then their result would be shifted down to $0.15 \pm 0.03~\mu$eV. This lower value favours the present lower strength, albeit not in a particularly statistically significant manner.


\subsection{\label{sec:dircap}Direct-Capture Yield Measurements}

The direct capture \Nepg~ cross section was measured by Rolfs \textit{et al.} \cite{rolfs1975} and Gorres \textit{et al.} \cite{gorres1983DC} in the energy range of $500 \leqslant E_{c.m.} \leqslant 1700$. These energies are too large to be of direct astrophysical importance, but were extrapolated down to lower energies using a direct capture model \cite{rolfs1973DCmodel}. Based on these results, an effective S-factor of $S(E) = 62$ keV$\cdot$b was extracted. More recently, the direct capture data has been extended to lower energies by both the TUNL \cite{Kelly-PRC95-2017} and LUNA \cite{Ferraro-PRL-2018} groups. Here we present data in the energy range of $282 \leqslant E_{c.m.} \leqslant 511$ keV. 

It is worth noting that carrying out direct capture measurements using inverse kinematics methods, such as  employed at DRAGON, means that the yield scales with the total non-resonant cross section, rather than the sum of partial cross sections for observed transitions. However, since we could only extract results from coincidence data, there is a second-order dependence on how unobserved transitions may impact the simulated coincidence efficiency. In order to obtain the primary branching ratios required for the simulation input, we extrapolated (to each measured C.M. energy) the partial cross sections predicted for each contributing state using the direct capture model of Ref \cite{rolfs1973DCmodel}, and proton spectroscopic factors published by Gorres \textit{et al.} \cite{gorres1983DC}. An approximate uncertainty of $40\%$ was assumed in these predictions, based on the recommendation from Hale \textit{et al.} \cite{Hale2002}. To understand how this might influence the coincidence efficiency, the extrapolated partial cross sections were randomised by folding in with a random Gaussian distribution with a sigma-width equal to $40\%$ of the central value. This procedure generated many possible primary branching ratio inputs, which were all simulated to obtain the spread in coincidence efficiencies one would expect based on the assumed uncertainty in the primary branch inputs. However, after some 30 simulations at each energy, the spread in calculated efficiencies turned out to be much less than the $10\%$ assumed systematic uncertainty in the simulation. The simulation input was further modified such that reactions are generated uniformly across the length of the target, in-keeping with a uniform cross-section arising due to non-resonant capture. For reference, the simulation normally generates reactions sampled from a Breit-Wigner shaped cross section; this would clearly be inappropriate for non-resonant capture, as systematic effects related to the recoil cone angle, energy spread, and BGO efficiency dependence on the origin of the reaction vertex, would not be properly reproduced.   

\begin{figure}[ht!]
 \includegraphics[width=0.45\textwidth]{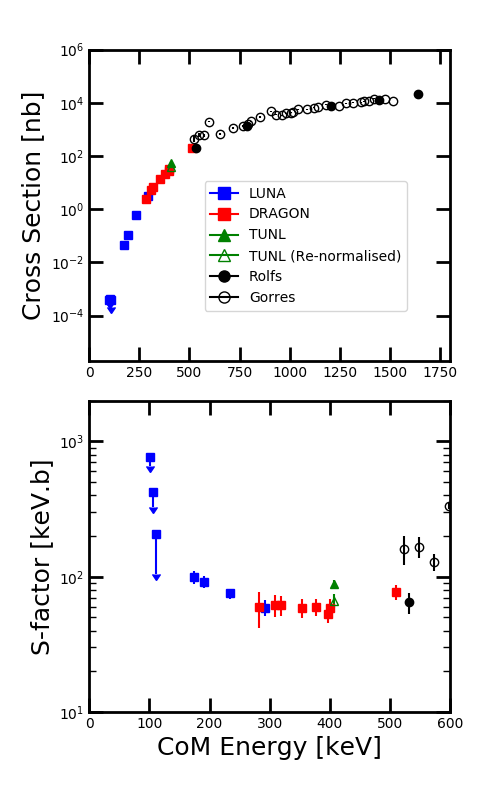}
\caption{Plot showing the direct capture cross section (upper panel) and astrophysical S-factor (lower panel) obtained from various data-sets. The TUNL value re-normalised to the strength of the 458-keV resonance from the present work is also plotted.} \label{fig:Sfactor}
 \end{figure}

\begin{table}[h!]
\caption{Direct capture cross sections and astrophysical S-factors determined from the present work} \smallskip
\centering
\begin{tabular}{l c c}
	\toprule[1pt]\midrule[0.3pt]
	\noalign{\smallskip}
	 $E_{c.m.}$ (keV)  & Cross Section (nb) & S-factor (keV.b) \\ \cline{1-3}
	\noalign{\smallskip}
	511(6.5) & $190.41 \pm 24.7$ & $78.0 \pm 10.3$ \\
	400(5.6) & $31.7 \pm 4.9$ & $60.0 \pm 9.8$ \\
	397(8.1) & $26.8 \pm 3.9$ & $54.9 \pm 8.4$ \\
	377(8.5) & $21.4 \pm 3.1$ & $61.2 \pm 9.4$ \\
	353(7.4) & $13.2 \pm 2.1$ & $59.2 \pm 9.6$ \\
	319(9.7) & $6.6 \pm 1.1$ & $56.6 \pm 10.0$ \\
	309(7.9) & $5.1 \pm 1.0$ & $55.6 \pm 10.7$ \\
	282(7.9) & $2.4 \pm 0.7$ & $50.8 \pm 15.0$ \\
	\noalign{\smallskip} 
	\midrule[0.3pt]\bottomrule[1pt]
\end{tabular}
\label{table:results_directcapture}
\end{table}

The resulting cross sections and astrophysical S-factors were calculated for each measured energy; these are plotted alongside literature data-sets in Figure \ref{fig:Sfactor}. From our results we find an astrophysical S-factor consistent with the previously adopted value of $62~\mathrm{keV}\cdot\mathrm{b}$. Unfortunately, our measurements do not extend down enough in energy to confirm the rise in the astrophysical S-factor seen by Ferraro \textit{et al.}, which the authors attribute to contributions from a broad sub-threshold resonance at $E=-130$ keV, arising due to a $J^{\pi} = 1/2^{+}$ state at $E_{x} = 8664$ keV.


\section{\label{sec:rate}Thermonuclear Reaction Rate}

In this work we report strength values for a total of seven resonances at center of mass energies of 149, 181, 248, 458, 610, 632 and 1222 keV. Since all of these are isolated narrow resonances, and there are no interference terms to consider, the total rate at a given temperature is calculated by summing the contribution of each resonance. The direct capture cross section was also measured, in the range of $282 \leqslant E_{c.m.} \leqslant 511$ keV, from which we derive an astrophysical S-factor of 60 keV.b. The thermonuclear rate, given in table \ref{table:reactionrate} of Appendix \ref{sec:reactionrate}, was calculating using the monte-carlo reaction rate calculator \texttt{RatesMC}. \texttt{RatesMC} computes the log-normal parameters describing the reaction rate at a given temperature. For a more detailed description of \texttt{RatesMC} the reader is referred to Ref \cite{iliadis2015}.

\begin{figure}[h!]
 \includegraphics[width=0.45\textwidth]{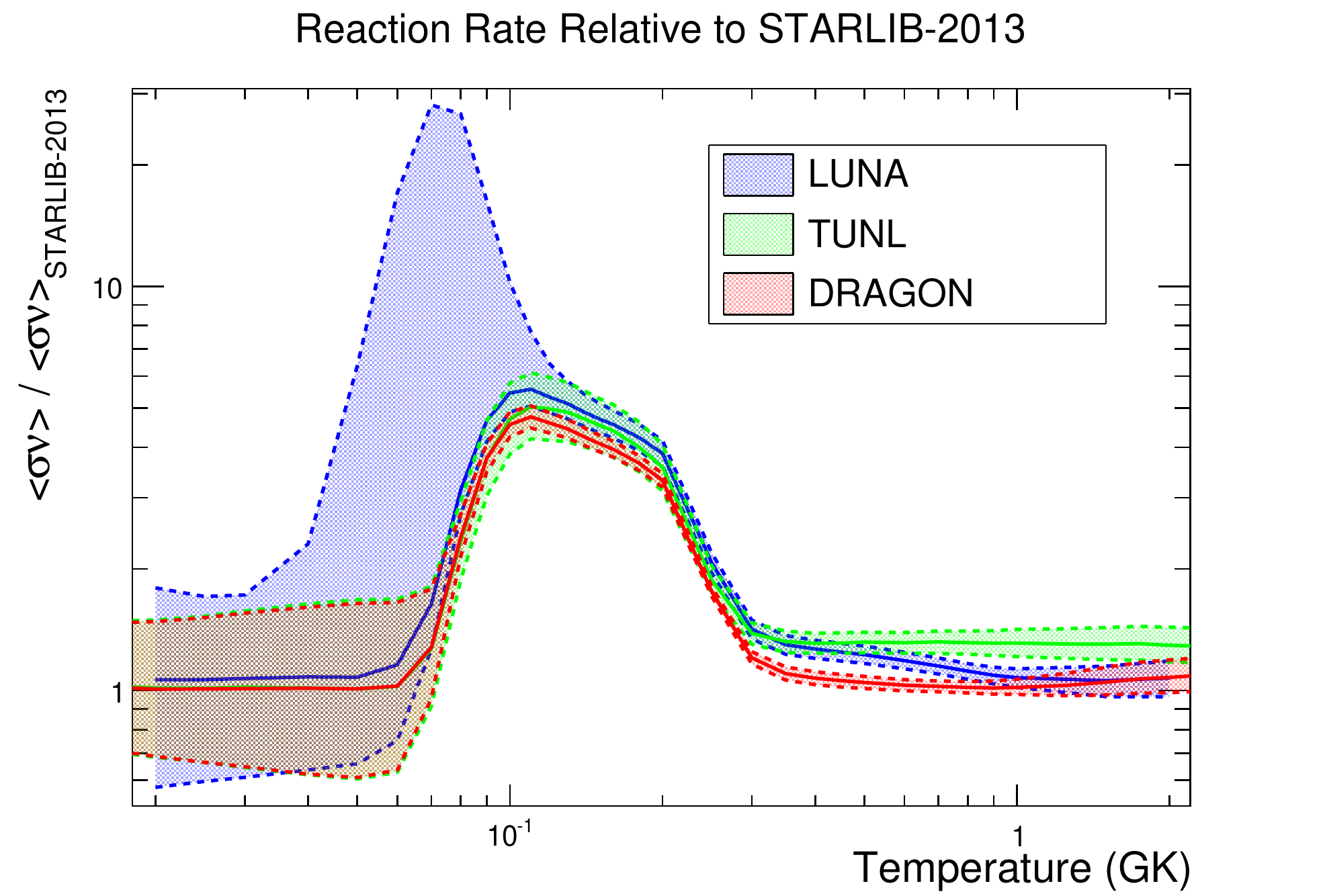}
\caption{Plot showing the LUNA, TUNL and present $^{22}$Ne$(p,\gamma)^{23}$Na reaction rates as a function of temperature in GK, expressed as a fraction of the STARLIB-2013 rate \cite{Sallaska-TAJSS207-2013}. The shaded regions represent the $1\sigma$ uncertainty bands associated with each rate.} \label{fig:reactionrate}
 \end{figure}

A comparison between the present rate and those put forward by the LUNA and TUNL groups is presented in figure \ref{fig:reactionrate}, expressed as a ratio over the STARLIB-2013 rate. The present rate is a factor of 4 higher than the STARLIB-2013 rate, following closely with the latest results from the LUNA and TUNL groups. The upper limit of the LUNA rate is discrepant with the present rate, since we do not take into account possible contribution from a tentative resonance at 68 keV.


\section{\label{sec:results}Astrophysical Impact}

\subsection{\label{sec:nova_calcs}Classical Novae}

The impact of the present rate was assessed for a variety of classical nova models, including carbon-oxygen (CO) and oxygen-neon (ONe) novae, with a range of considered white dwarf masses. These were modelled using the one-dimensional, spherically symmetric, implicit hydro-dynamical code SHIVA \cite{jose1998,jose2016}, which has been used extensively to model nucleosynthesis in classical novae. 

Final abundances of nuclei in the Ne-Al region, calculated assuming the present $^{22}$Ne$(p,\gamma)^{23}$Na rate and the previous STARLIB-2013 rate \cite{sallaska2013}, are tabulated in Appendix \ref{sec:novae_calcs}. These simulations show that the most wide-spread changes in the ejecta abundances occur for the $1.15~M_{\odot}$ CO nova model, which exhibits changes of more than $10\%$ for $^{20}$Ne, $^{21}$Ne, $^{22}$Ne, $^{22}$Na, $^{23}$Na, $^{25}$Mg, $^{26}$Mg, $^{26}$Al, and $^{27}$Al. The most significant abundance change of any single isotope was $^{23}$Na, with approximately a factor of 2 enhancement for both CO nova models. For the ONe nova models, the $^{22}$Ne content is reduced by almost a factor of 2 in both cases, while only modest changes are predicted for all other isotopes considered, with the exception of $^{24}$Mg which is enhanced by $\sim15$\% in the 1.25 $M_{\odot}$ ONe nova model. 

The magnesium isotopic ratios $^{25}$Mg/$^{24}$Mg and $^{26}$Mg/$^{25}$Mg warrant closer inspection. These ratios have been studied as a possible means of identifying pre-solar grains of putative classical nova origin, and to provide model constraints on important model parameters such as the peak temperature achieved during the outburst \cite{jose2004}. In the case of CO novae, synthesis of Mg is very sensitive to the peak temperature reached, and hence the underlying WD mass \cite{jose2004}. The sensitivity study performed by Iliadis \textit{et al.} \cite{iliadis-AJS142-2002} showed that the predicted final abundances of $^{24}$Mg and $^{25}$Mg for the 1.15 M$_{\odot}$ CO nova model change by up to a factor of 5, as a result of varying the $^{22}$Ne$(p,\gamma)^{23}$Na rate within its prior uncertainties. The newly determined rate drastically limits the reaction rate uncertainty in the relevant temperature range ($T_{\textrm{peak}}= 170$ MK). Indeed, by varying the current rate within its respective low and high uncertainty limits, changes of less than 7\% are observed for all the Mg isotope mass fractions.

Furthermore, the new rate seems to accentuate differences in the Mg isotope ratios between the 1.0 M$_{\odot}$ and 1.15 M$_{\odot}$ models. In comparison to the STARLIB-2013 rate, the calculations performed with the new $^{22}$Ne$(p,\gamma)^{23}$Na rate result in a 24$\%$ increase and a 13$\%$ decrease in the $^{25}$Mg/$^{24}$Mg and $^{26}$Mg/$^{25}$Mg isotopic ratios, respectively, for the 1.15 M$_{\odot}$ model. However, no significant change is seen for the Mg isotopes in the 1.0 M$_{\odot}$ model. This result could be of potential interest for using Mg isotopic ratios in pre-solar grains as a thermometer for the peak temperatures reached during the outburst. Further work should be undertaken to reassess the sensitivity of magnesium isotopic ratios in CO novae to current nuclear reaction rate uncertainties in the Ne-Al region, incorporating the new $^{22}$Ne$(p,\gamma)^{23}$Na rate and associated uncertainties.

Enhanced neon content in meteoritic samples has historically been proposed as a fingerprint for identifying pre-solar grains of classical nova origin, particularly in terms of excess $^{22}$Ne content associated with the decay of $^{22}$Na \cite{clayton1975}. The $^{20}$Ne/$^{22}$Ne isotopic ratio is also of interest for distinguishing between CO and ONe novae; the latter are expected to have very large ratios of $^{20}$Ne/$^{22}$Ne $>100$, whereas CO novae models yield ratios of $^{20}$Ne/$^{22}$Ne $<1$ \cite{jose2004}. The present rate leads to more efficient destruction of $^{22}$Ne by approximately a factor of 2 over the previous rate, while leaving the mass fraction of $^{22}$Na released in the ejecta completely untouched. The previously assumed uncertainty in Neon abundances, due to the $^{22}$Ne$(p,\gamma)^{23}$Na rate, is also drastically reduced from orders of magnitude to a few percent, marking a significant improvement in the nuclear physics input uncertainties related to key isotopic ratios predicted for classical nova nucleosynthesis. 



\subsection{\label{sec:agb_stars}AGB Stars}


The rate calculated through this work was implemented in a series of nucleosythesis network calculations performed using the NuGrid multi-zone post-processing code \texttt{MPPNP} \cite{ritter2018}. Three stellar models were considered for this work, each generated using the stellar evolution code \texttt{MESA} \cite{paxton2010} and evolved up to the AGB phase. These models also include a recently developed treatment for convective boundary mixing occurring at the bottom of convective envelope during third dredge-up \cite{battino2016}.

The $5M_{\odot}$ $(z = 0.006)$ model was used to assess the impact of the present rate, in comparison to the STARLIB-2013 rate \cite{sallaska2013}, for hot bottom burning in thermally pulsing AGB stars. In addition, simulations of low mass AGB stars were performed to assess the impact of the present rate on the formation of the so-called sodium pocket\cite{Cristallo-APJ696-2009,Lucatello-APJ-2011}. In low mass AGB stars of solar metallicity, recent stellar models predict that the sodium pocket should be a major source of $^{23}$Na, with production of $^{23}$Na thought to be related to ingestion of the sodium pocket during third dredge-up \cite{Cristallo-APJ696-2009}.

\begin{figure}[h!]
 \includegraphics[width=0.45\textwidth]{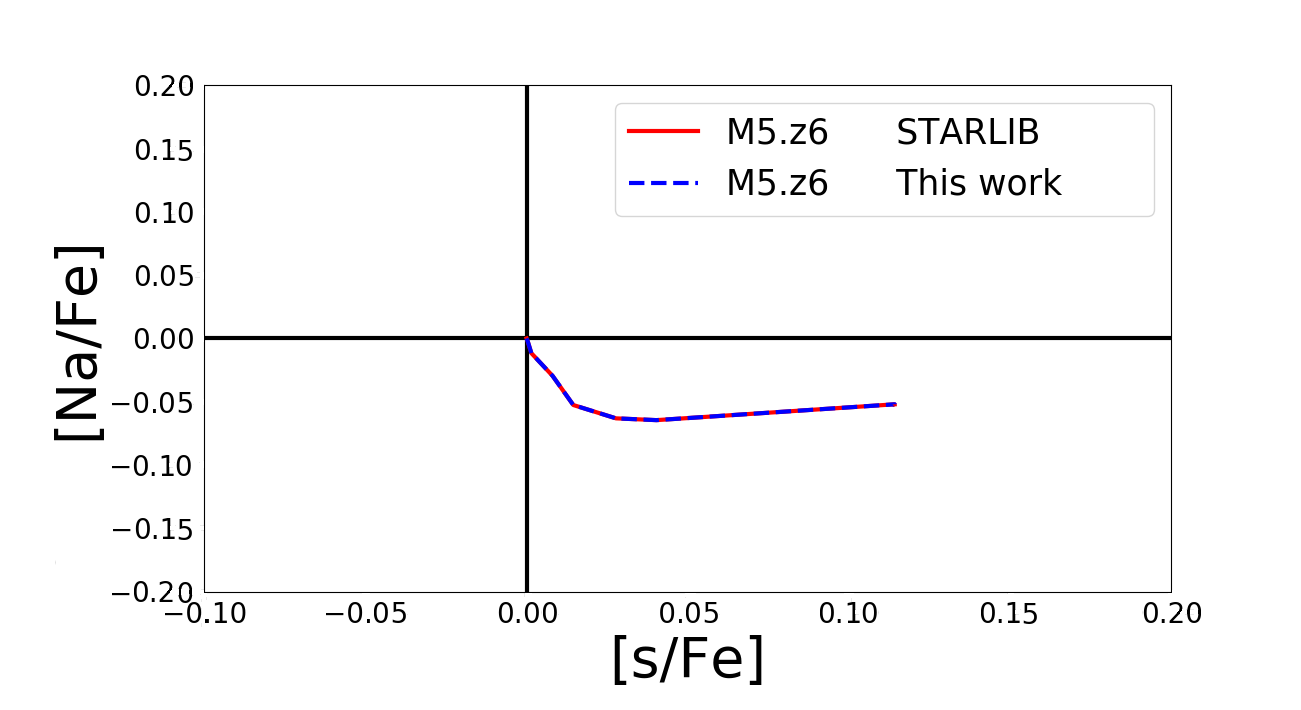}
\caption{Predicted surface [Na/Fe] abundance ratio plotted as a function of s-process
element abundances [s/Fe] for a $5M_{\odot}$ ($z = 0.006$) AGB star.} \label{fig:agb5msol}
 \end{figure}

Despite a factor of 4 enhancement at $T=100$ MK over the previous thermonuclear rate, there appears to be very little impact on $^{23}$Na production during HBB in the $5M_{\odot}$ TP-AGB star model, as demonstrated in Figure \ref{fig:agb5msol}. This is in contrast with the significant enhancement (factor $\sim3$) obtained from similar calculations using the LUNA rate, which was investigated by Slemer \textit{et al.} \cite{slemer2016}. This is likely a consequence of two factors. The first factor arises due to the significant enhancement at 100 MK seen in the $^{22}$Ne$(p,\gamma)^{23}$Na rate put forward by Cavanna et al. \cite{Cavanna-PRL115-2015}, due largely to their treatment of tentative resonances at 68 and 100 keV. The second results from the models performed by Slemmer \textit{et al.} \cite{slemer2016} not taking into account neutron capture reactions. However, neglecting neutron capture reactions is potentially problematic, given the results of Cristallo \textit{et al.} \cite{cristallo2006}, which show that in low-metalicity $(z=10^{-4})$ AGB stellar models neutron capture on $^{22}$Ne can contribute $13\%$ and $35\%$ of the total surface $^{23}$Na abundance from $^{13}$C$(\alpha,n)$ and $^{22}$Ne$(\alpha,n)$ burning respectively. The model calculations presented in this work include neutron capture reactions.


\begin{figure}[h!]
 \includegraphics[width=0.45\textwidth]{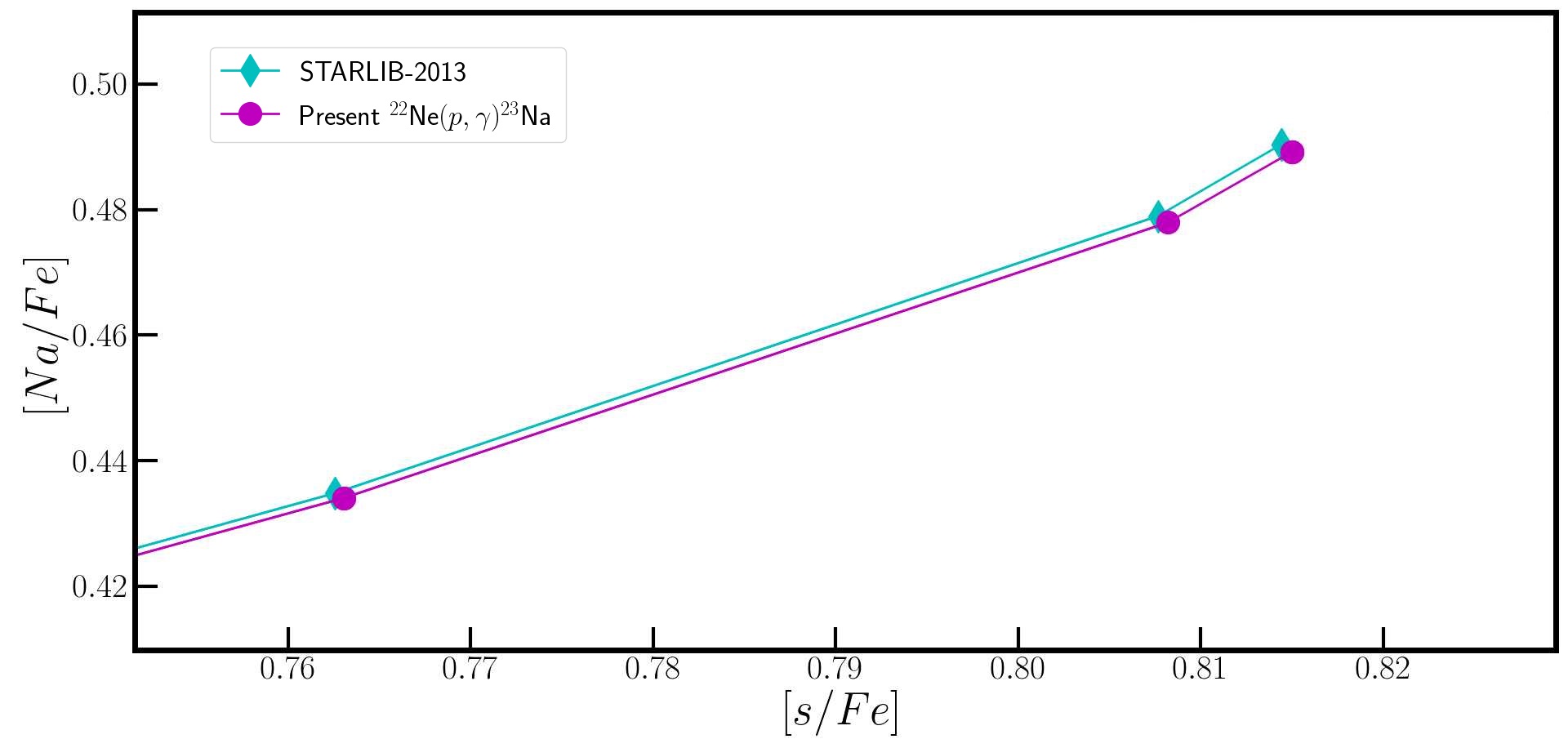}
\caption{Predicted surface [Na/Fe] abundance ratio plotted as a function of Sprocess
element abundances [s/Fe] for a $2M_{\odot}$ $(z = 0.006)$ AGB star.} \label{fig:agb2msol}
 \end{figure}

In the case of low mass AGB stars, formation of the sodium pocket in also appears to be negligibly affected by adopting the present $^{22}$Ne$(p,\gamma)^{23}$Na rate. The resulting small effect on the surface [Na/Fe] ratio is shown by Figure \ref{fig:agb2msol}. No discernible changes in the surface Na abundance could be seen for the lower metallicity (Z=0.001) model.


\section{Conclusions}

In summary, the $^{22}$Ne$(p,\gamma)^{23}$Na reaction has, for the first time, been investigated directly in inverse kinematics. As such, the present work is subject to different experimental systematics than previous studies already found in the literature. A total of 7 resonances were measured, located at center of mass energies: 149, 181, 248, 458, 610, 632 and 1222 keV.

The important reference resonance at 458 keV was measured to have a strength value of $\omega\gamma_{458} = 0.44 \pm 0.02$ eV. This is significantly lower than values published in two recent studies \cite{Kelly-PRC92-2015,Depalo-PRC92-2015}. In the case of the three lowest energy resonances, which have the strongest influence on the reaction rate at stellar temperatures, we find close agreement with recent studies conducted at LUNA \cite{Cavanna-PRL115-2015,Cavanna-PRL120E-2018,Depalo-PRC94-2016,Ferraro-PRL-2018} and TUNL \cite{Kelly-PRC95-2017}.

The non-resonant contribution to the $^{22}$Ne$(p,\gamma)^{23}$Na reaction rate was also measured, in the energy range of $282 \leqslant E_{c.m.} \leqslant 511$ keV. The astrophysical s-factor associated with direct capture is found to be consistent with the previous work of Rolfs \textit{et al.} \cite{rolfs1975}. Reported Erickson fluctuations in the direct capture cross section observed by G\"{o}rres \textit{et al.} \cite{gorres1983DC} were not found to persist in the energy range considered here. Unfortunately, the data points contributed by the present study do not extend low enough in energy to observe the influence of the $E_{x} = 8664$ keV sub-threshold state, which results in the upturn in the astrophysical s-factor observed in the most recent LUNA study \cite{Ferraro-PRL-2018}.

Our newly proposed rate follows closely with that put forward by the TUNL group. The key difference with respect to the rate published by the LUNA group is a consequence of their inclusion of upper limits from tentative resonances at 68 and 100 keV. The associated states, tentatively observed by Powers \textit{et al.} \cite{Powers71}, have not been observed in a subsequent $(^{3}$He,$d)$ transfer study \cite{Hale2002}, nor in the unselective $(p,p\prime)$ reaction study by Moss \textit{et al.} \cite{moss1976}. These states have thus been neglected by the present work, as well as by Kelly \textit{et al.} \cite{Kelly-PRC95-2017}. Furthermore, preliminary analysis of a high resolution $^{23}$Na$(p,p\prime)^{23}$Na study conducted at the Munich Maier-Leibnitz Laboratory shows no signal above background in the relevant excitation region. Details of this study will be put forward in a forthcoming publication.

The impact of our newly proposed rate was assessed for both classical nova and AGB star nucleosynthesis. As a consequence of the present work, uncertainties in the predicted ejecta abundances in the Ne-Al region from CO and ONe novae have been drastically reduced. No significant enhancement in $^{23}$Na production is evident in the $M=$ 2, 3, 5 $M_{\odot}$ AGB star models considered in this work. The $^{22}$Ne$(p,\gamma)$ rate is now sufficiently well constrained in the major astrophysical environment thought to contribute to the Na-O anti-correlation.

\section{Acknowledgements}

The authors thank the ISAC operations and technical staff at TRIUMF. TRIUMF’s core operations are supported via a contribution from the federal government through the National Research Council Canada, and the Government of British Columbia provides building capital funds. DRAGON is supported by funds from the National Sciences and Engineering Research Council of Canada. UK authors gratefully acknowledge support from the Science and Technology Facilities Council (STFC). J. Jos\'e acknowledges support from the Spanish MINECO grant AYA2017\hyph{}86274\hyph{}P, the EU FEDER funds and the AGAUR/ Generalitat de Catalunya grant SGR-661/2017. Authors from the Colorado School of Mines acknowledge funding via the U.S. Department of Energy grant DE\hyph{}FG02\hyph{}93ER40789. U. Battino acknowledges support from the European Research Council (ERC\hyph{}2015\hyph{}STG Nr.677497). This article also benefited from discussions within the ‘ChETEC’ COST Action (CA16117). The authors also thank R. Longland for his support in calculating the thermonuclear reaction rate presented in this work.

\bibliography{Neon22references}

\appendix

\section{Thermonuclear Reaction Rate}
\label{sec:reactionrate}

This appendix contains the total thermonuclear reaction rate adopted following this work. The thermonuclear rate was computed using the \texttt{RatesMC} code, which calculates the log-normal parameters $\mu$ and $\sigma$ describing the reaction rate at a given temperature. The column labelled `A-D statistic' refers to the Anderson-Darling statistic, indicating how well a log-normal distribution describes the rate at a given temperature. An A-D statistic of less than $\approx1$ indicates that the rate is well described by a log-normal distribution. However, it has been shown that the assumption of a log-normal distributed reaction rate holds for A-D statistics in the $\approx 1-30$ range \cite{Iliadis-NPA841-2010}.

{\setlength{\tabcolsep}{8pt}

\begin{longtable*}{c c c c c c c}

\caption{Tabulated $^{22}$Ne$(p,\gamma)
^{23}$Na total thermonuclear reaction rate determined from the present work, expressed in units of cm$^{3}$ mol$^{-1}$ s$^{-1}$.} 

\label{table:reactionrate} \\

\toprule[1pt]\midrule[0.3pt]

 T [GK] & Low rate & Medium rate & High rate & Log-normal $\mu$ & Log-normal $\sigma$ & A-D statistic \\ \hline \noalign{\smallskip}
\endfirsthead

\multicolumn{7}{c}%
{{\tablename\ \thetable{} -- \textit{Continued from previous page}}} \\
\toprule[1pt]\midrule[0.3pt]
T [GK] & Low rate & Medium rate & High rate & Log-normal $\mu$ & Log-normal $\sigma$ & A-D statistic \\ \hline \noalign{\smallskip}
\endhead

\hline \noalign{\smallskip} \multicolumn{7}{c}{{\textit{Continued on next page}}} \\ \noalign{\smallskip} \hline
\endfoot

\midrule[0.3pt]\bottomrule[1pt]
\endlastfoot

        0.010 & 4.21$\times$10$^{-25}$ & 6.75$\times$10$^{-25}$ &
      1.08$\times$10$^{-24}$ &  -5.566$\times$10$^{+01}$ &
       4.83$\times$10$^{-01}$  & 7.88$\times$10$^{-01}$ \\ 
0.011 & 1.58$\times$10$^{-23}$ & 2.44$\times$10$^{-23}$ &
      3.78$\times$10$^{-23}$ &  -5.207$\times$10$^{+01}$ &
       4.39$\times$10$^{-01}$  & 5.26$\times$10$^{-01}$ \\ 
0.012 & 3.21$\times$10$^{-22}$ & 4.81$\times$10$^{-22}$ &
      7.21$\times$10$^{-22}$ &  -4.909$\times$10$^{+01}$ &
       4.09$\times$10$^{-01}$  & 2.75$\times$10$^{-01}$ \\ 
0.013 & 4.03$\times$10$^{-21}$ & 5.93$\times$10$^{-21}$ &
      8.71$\times$10$^{-21}$ &  -4.657$\times$10$^{+01}$ &
       3.89$\times$10$^{-01}$  & 2.04$\times$10$^{-01}$ \\ 
0.014 & 3.49$\times$10$^{-20}$ & 5.08$\times$10$^{-20}$ &
      7.37$\times$10$^{-20}$ &  -4.443$\times$10$^{+01}$ &
       3.78$\times$10$^{-01}$  & 1.50$\times$10$^{-01}$ \\ 
0.015 & 2.23$\times$10$^{-19}$ & 3.24$\times$10$^{-19}$ &
      4.68$\times$10$^{-19}$ &  -4.257$\times$10$^{+01}$ &
       3.72$\times$10$^{-01}$  & 1.61$\times$10$^{-01}$ \\ 
0.016 & 1.13$\times$10$^{-18}$ & 1.62$\times$10$^{-18}$ &
      2.35$\times$10$^{-18}$ &  -4.096$\times$10$^{+01}$ &
       3.70$\times$10$^{-01}$  & 1.22$\times$10$^{-01}$ \\ 
0.018 & 1.63$\times$10$^{-17}$ & 2.36$\times$10$^{-17}$ &
      3.44$\times$10$^{-17}$ &  -3.828$\times$10$^{+01}$ &
       3.74$\times$10$^{-01}$  & 1.26$\times$10$^{-01}$ \\ 
0.020 & 1.35$\times$10$^{-16}$ & 1.98$\times$10$^{-16}$ &
      2.90$\times$10$^{-16}$ &  -3.616$\times$10$^{+01}$ &
       3.83$\times$10$^{-01}$  & 3.09$\times$10$^{-01}$ \\ 
0.025 & 5.69$\times$10$^{-15}$ & 8.64$\times$10$^{-15}$ &
      1.30$\times$10$^{-14}$ &  -3.238$\times$10$^{+01}$ &
       4.12$\times$10$^{-01}$  & 6.77$\times$10$^{-01}$ \\ 
0.030 & 6.55$\times$10$^{-14}$ & 1.02$\times$10$^{-13}$ &
      1.57$\times$10$^{-13}$ &  -2.992$\times$10$^{+01}$ &
       4.39$\times$10$^{-01}$  & 7.32$\times$10$^{-01}$ \\ 
0.040 & 1.25$\times$10$^{-12}$ & 2.05$\times$10$^{-12}$ &
      3.25$\times$10$^{-12}$ &  -2.692$\times$10$^{+01}$ &
       4.78$\times$10$^{-01}$  & 8.10$\times$10$^{-01}$ \\ 
0.050 & 7.00$\times$10$^{-12}$ & 1.16$\times$10$^{-11}$ &
      1.89$\times$10$^{-11}$ &  -2.519$\times$10$^{+01}$ &
       4.99$\times$10$^{-01}$  & 6.64$\times$10$^{-01}$ \\ 
0.060 & 2.36$\times$10$^{-11}$ & 3.81$\times$10$^{-11}$ &
      6.13$\times$10$^{-11}$ &  -2.399$\times$10$^{+01}$ &
       4.79$\times$10$^{-01}$  & 6.58$\times$10$^{-01}$ \\ 
0.070 & 1.00$\times$10$^{-10}$ & 1.34$\times$10$^{-10}$ &
      1.87$\times$10$^{-10}$ &  -2.271$\times$10$^{+01}$ &
       3.17$\times$10$^{-01}$  & 2.73$\times$10$^{+01}$ \\ 
0.080 & 8.25$\times$10$^{-10}$ & 9.22$\times$10$^{-10}$ &
      1.05$\times$10$^{-09}$ &  -2.080$\times$10$^{+01}$ &
       1.26$\times$10$^{-01}$  & 2.90$\times$10$^{+01}$ \\ 
0.090 & 6.57$\times$10$^{-09}$ & 7.10$\times$10$^{-09}$ &
      7.71$\times$10$^{-09}$ &  -1.876$\times$10$^{+01}$ &
       8.12$\times$10$^{-02}$  & 1.23$\times$10$^{+00}$ \\ 
0.100 & 3.90$\times$10$^{-08}$ & 4.18$\times$10$^{-08}$ &
      4.49$\times$10$^{-08}$ &  -1.699$\times$10$^{+01}$ &
       7.10$\times$10$^{-02}$  & 3.04$\times$10$^{-01}$ \\ 
0.110 & 1.74$\times$10$^{-07}$ & 1.86$\times$10$^{-07}$ &
      1.98$\times$10$^{-07}$ &  -1.550$\times$10$^{+01}$ &
       6.38$\times$10$^{-02}$  & 1.97$\times$10$^{-01}$ \\ 
0.120 & 6.20$\times$10$^{-07}$ & 6.56$\times$10$^{-07}$ &
      6.95$\times$10$^{-07}$ &  -1.424$\times$10$^{+01}$ &
       5.81$\times$10$^{-02}$  & 1.56$\times$10$^{-01}$ \\ 
0.130 & 1.83$\times$10$^{-06}$ & 1.93$\times$10$^{-06}$ &
      2.03$\times$10$^{-06}$ &  -1.316$\times$10$^{+01}$ &
       5.37$\times$10$^{-02}$  & 3.04$\times$10$^{-01}$ \\ 
0.140 & 4.65$\times$10$^{-06}$ & 4.88$\times$10$^{-06}$ &
      5.14$\times$10$^{-06}$ &  -1.223$\times$10$^{+01}$ &
       5.04$\times$10$^{-02}$  & 5.66$\times$10$^{-01}$ \\ 
0.150 & 1.05$\times$10$^{-05}$ & 1.10$\times$10$^{-05}$ &
      1.15$\times$10$^{-05}$ &  -1.142$\times$10$^{+01}$ &
       4.78$\times$10$^{-02}$  & 6.98$\times$10$^{-01}$ \\ 
0.160 & 2.14$\times$10$^{-05}$ & 2.24$\times$10$^{-05}$ &
      2.35$\times$10$^{-05}$ &  -1.071$\times$10$^{+01}$ &
       4.58$\times$10$^{-02}$  & 7.12$\times$10$^{-01}$ \\ 
0.180 & 7.14$\times$10$^{-05}$ & 7.44$\times$10$^{-05}$ &
      7.77$\times$10$^{-05}$ &  -9.505$\times$10$^{+00}$ &
       4.24$\times$10$^{-02}$  & 6.51$\times$10$^{-01}$ \\ 
0.200 & 1.93$\times$10$^{-04}$ & 2.01$\times$10$^{-04}$ &
      2.09$\times$10$^{-04}$ &  -8.514$\times$10$^{+00}$ &
       3.89$\times$10$^{-02}$  & 4.86$\times$10$^{-01}$ \\ 
0.250 & 1.83$\times$10$^{-03}$ & 1.88$\times$10$^{-03}$ &
      1.94$\times$10$^{-03}$ &  -6.276$\times$10$^{+00}$ &
       2.85$\times$10$^{-02}$  & 2.51$\times$10$^{-01}$ \\ 
0.300 & 2.00$\times$10$^{-02}$ & 2.07$\times$10$^{-02}$ &
      2.15$\times$10$^{-02}$ &  -3.876$\times$10$^{+00}$ &
       3.56$\times$10$^{-02}$  & 5.13$\times$10$^{-01}$ \\ 
0.350 & 1.58$\times$10$^{-01}$ & 1.64$\times$10$^{-01}$ &
      1.70$\times$10$^{-01}$ &  -1.807$\times$10$^{+00}$ &
       3.80$\times$10$^{-02}$  & 3.95$\times$10$^{-01}$ \\ 
0.400 & 7.98$\times$10$^{-01}$ & 8.28$\times$10$^{-01}$ &
      8.59$\times$10$^{-01}$ &  -1.884$\times$10$^{-01}$ &
       3.73$\times$10$^{-02}$  & 3.87$\times$10$^{-01}$ \\ 
0.450 & 2.86$\times$10$^{+00}$ & 2.96$\times$10$^{+00}$ &
      3.07$\times$10$^{+00}$ &  1.086$\times$10$^{+00}$ &
       3.60$\times$10$^{-02}$  & 4.59$\times$10$^{-01}$ \\ 
0.500 & 7.98$\times$10$^{+00}$ & 8.26$\times$10$^{+00}$ &
      8.55$\times$10$^{+00}$ &  2.112$\times$10$^{+00}$ &
       3.45$\times$10$^{-02}$  & 5.08$\times$10$^{-01}$ \\ 
0.600 & 3.79$\times$10$^{+01}$ & 3.92$\times$10$^{+01}$ &
      4.04$\times$10$^{+01}$ &  3.668$\times$10$^{+00}$ &
       3.19$\times$10$^{-02}$  & 5.85$\times$10$^{-01}$ \\ 
0.700 & 1.18$\times$10$^{+02}$ & 1.22$\times$10$^{+02}$ &
      1.26$\times$10$^{+02}$ &  4.803$\times$10$^{+00}$ &
       3.08$\times$10$^{-02}$  & 8.30$\times$10$^{-01}$ \\ 
0.800 & 2.83$\times$10$^{+02}$ & 2.92$\times$10$^{+02}$ &
      3.02$\times$10$^{+02}$ &  5.678$\times$10$^{+00}$ &
       3.28$\times$10$^{-02}$  & 2.73$\times$10$^{+00}$ \\ 
0.900 & 5.68$\times$10$^{+02}$ & 5.89$\times$10$^{+02}$ &
      6.12$\times$10$^{+02}$ &  6.380$\times$10$^{+00}$ &
       3.82$\times$10$^{-02}$  & 9.77$\times$10$^{+00}$ \\ 
1.000 & 1.01$\times$10$^{+03}$ & 1.05$\times$10$^{+03}$ &
      1.10$\times$10$^{+03}$ &  6.959$\times$10$^{+00}$ &
       4.59$\times$10$^{-02}$  & 1.88$\times$10$^{+01}$ \\ 
1.250 & 2.94$\times$10$^{+03}$ & 3.12$\times$10$^{+03}$ &
      3.34$\times$10$^{+03}$ &  8.051$\times$10$^{+00}$ &
       6.63$\times$10$^{-02}$  & 2.74$\times$10$^{+01}$ \\ 
1.500 & 6.25$\times$10$^{+03}$ & 6.72$\times$10$^{+03}$ &
      7.32$\times$10$^{+03}$ &  8.820$\times$10$^{+00}$ &
       8.16$\times$10$^{-02}$  & 2.53$\times$10$^{+01}$ \\ 
1.750 & 1.09$\times$10$^{+04}$ & 1.19$\times$10$^{+04}$ &
      1.30$\times$10$^{+04}$ &  9.388$\times$10$^{+00}$ &
       9.09$\times$10$^{-02}$  & 2.25$\times$10$^{+01}$ \\ 
2.000 & 1.68$\times$10$^{+04}$ & 1.83$\times$10$^{+04}$ &
      2.02$\times$10$^{+04}$ &  9.822$\times$10$^{+00}$ &
       9.58$\times$10$^{-02}$  & 2.04$\times$10$^{+01}$ \\ 
2.500 & 3.10$\times$10$^{+04}$ & 3.39$\times$10$^{+04}$ &
      3.75$\times$10$^{+04}$ &  1.044$\times$10$^{+01}$ &
       9.78$\times$10$^{-02}$  & 1.82$\times$10$^{+01}$ \\ 
3.000 & 4.66$\times$10$^{+04}$ & 5.10$\times$10$^{+04}$ &
      5.63$\times$10$^{+04}$ &  1.085$\times$10$^{+01}$ &
       9.51$\times$10$^{-02}$  & 1.75$\times$10$^{+01}$ \\ 
3.500 & 6.22$\times$10$^{+04}$ & 6.77$\times$10$^{+04}$ &
      7.45$\times$10$^{+04}$ &  1.113$\times$10$^{+01}$ &
       9.11$\times$10$^{-02}$  & 1.71$\times$10$^{+01}$ \\ 
4.000 & 7.64$\times$10$^{+04}$ & 8.29$\times$10$^{+04}$ &
      9.08$\times$10$^{+04}$ &  1.133$\times$10$^{+01}$ &
       8.70$\times$10$^{-02}$  & 1.68$\times$10$^{+01}$ \\ 
5.000 & 9.95$\times$10$^{+04}$ & 1.07$\times$10$^{+05}$ &
      1.17$\times$10$^{+05}$ &  1.159$\times$10$^{+01}$ &
       8.00$\times$10$^{-02}$  & 1.59$\times$10$^{+01}$ \\ 
6.000 & 1.15$\times$10$^{+05}$ & 1.23$\times$10$^{+05}$ &
      1.33$\times$10$^{+05}$ &  1.173$\times$10$^{+01}$ &
       7.48$\times$10$^{-02}$  & 1.46$\times$10$^{+01}$ \\ 
7.000 & 1.24$\times$10$^{+05}$ & 1.33$\times$10$^{+05}$ &
      1.43$\times$10$^{+05}$ &  1.180$\times$10$^{+01}$ &
       7.11$\times$10$^{-02}$  & 1.33$\times$10$^{+01}$ \\ 
8.000 & 1.29$\times$10$^{+05}$ & 1.38$\times$10$^{+05}$ &
      1.48$\times$10$^{+05}$ &  1.184$\times$10$^{+01}$ &
       6.83$\times$10$^{-02}$  & 1.21$\times$10$^{+01}$ \\ 
9.000 & 1.31$\times$10$^{+05}$ & 1.39$\times$10$^{+05}$ &
      1.49$\times$10$^{+05}$ &  1.184$\times$10$^{+01}$ &
       6.62$\times$10$^{-02}$  & 1.10$\times$10$^{+01}$ \\ 
10.000 & 1.30$\times$10$^{+05}$ & 1.38$\times$10$^{+05}$ &
      1.48$\times$10$^{+05}$ &  1.184$\times$10$^{+01}$ &
       6.46$\times$10$^{-02}$  & 1.00$\times$10$^{+01}$ \\ 
\end{longtable*}
}

\section{Classical Novae Model Calculations}
\label{sec:novae_calcs}

This appendix contains tables of isotope mass fractions in the Ne-Al mass range ejected assuming a variety of classical novae models. Two carbon-oxygen and two oxygen-neon novae models were considered, see text in section \ref{sec:nova_calcs} for a summary of key findings. The models were generated using the one dimensional spherically symmetric hydrodynamic code SHIVA \cite{jose1998,jose2016}.

\begin{table*}[htb!]
\centering
\caption{Predicted ejecta mass fractions for a 1.0M$_{\odot}$ CO nova model in the Ne-Al region. Total mass of the ejected envelope is $3.35 \times 10^{-5}$ M$_{\odot}$.}
\begin{tabular}{c c c c c}
	\toprule[1pt]\midrule[0.3pt] \noalign{\smallskip}
	Nuclide & STARLIB-2013 & Low Rate & Medium Rate & High Rate \\ \cmidrule[0.3pt]{1-5} \noalign{\smallskip}
$^{20}$Ne & $1.28 \times 10^{-3}$ & $1.37 \times 10^{-3}$ & $1.35 \times 10^{-3}$ & $1.34 \times 10^{-3}$ \\
$^{21}$Ne & $1.45 \times 10^{-7}$ & $1.53 \times 10^{-7}$ & $1.52 \times 10^{-7}$ & $1.51 \times 10^{-7}$ \\
$^{22}$Ne & $2.50 \times 10^{-3}$ & $2.32 \times 10^{-3}$ & $2.36 \times 10^{-3}$ & $2.39 \times 10^{-3}$ \\
$^{22}$Na & $6.97 \times 10^{-7}$ & $7.32 \times 10^{-7}$ & $7.26 \times 10^{-3}$ & $7.20 \times 10^{-7}$ \\
$^{23}$Na & $2.61 \times 10^{-5}$ & $6.58 \times 10^{-5}$ & $5.83 \times 10^{-5}$ & $5.22 \times 10^{-5}$ \\
$^{24}$Mg & $1.36 \times 10^{-5}$ & $1.44 \times 10^{-5}$ & $1.43 \times 10^{-5}$ & $1.43 \times 10^{-5}$ \\
$^{25}$Mg & $4.02 \times 10^{-4}$ & $4.39 \times 10^{-4}$ & $4.32 \times 10^{-4}$ & $4.26 \times 10^{-4}$ \\
$^{26}$Mg & $4.17 \times 10^{-5}$ & $4.22 \times 10^{-5}$ & $4.21 \times 10^{-5}$ & $4.20 \times 10^{-5}$ \\
$^{26}$Al & $3.25 \times 10^{-5}$ & $3.46 \times 10^{-5}$ & $3.42 \times 10^{-5}$ & $3.39 \times 10^{-5}$ \\
$^{27}$Al & $8.21 \times 10^{-5}$ & $8.33 \times 10^{-5}$ & $8.30 \times 10^{-5}$ & $8.29 \times 10^{-5}$ \\
		\midrule[0.3pt]\bottomrule[1pt] 
\end{tabular}
\label{table:CO1model}
\end{table*}

\begin{table*}[h!]
\centering
\caption{Predicted ejecta mass fractions for a 1.15 M$_{\odot}$ CO nova model in the Ne-Al region. Total mass of the ejected envelope is $1.44 \times 10^{-5}$ M$_{\odot}$.}
\begin{tabular}{c c c c c}
\toprule[1pt]\midrule[0.3pt] \noalign{\smallskip}
	Nuclide & STARLIB-2013 & Low Rate & Medium Rate & High Rate \\ \cmidrule[0.3pt]{1-5} \noalign{\smallskip}
$^{20}$Ne & $1.42 \times 10^{-3}$ & $1.67 \times 10^{-3}$ & $1.64 \times 10^{-3}$ & $1.60 \times 10^{-3}$ \\
$^{21}$Ne & $2.52 \times 10^{-7}$ & $2.96 \times 10^{-7}$ & $2.88 \times 10^{-7}$ & $2.82 \times 10^{-7}$ \\
$^{22}$Ne & $2.52 \times 10^{-3}$ & $1.80 \times 10^{-3}$ & $1.87 \times 10^{-3}$ & $1.83 \times 10^{-3}$ \\
$^{22}$Na & $7.52 \times 10^{-7}$ & $8.69 \times 10^{-7}$ & $8.50 \times 10^{-3}$ & $8.34 \times 10^{-7}$ \\
$^{23}$Na & $1.73 \times 10^{-5}$ & $3.61 \times 10^{-5}$ & $3.32 \times 10^{-5}$ & $3.07 \times 10^{-5}$ \\
$^{24}$Mg & $6.13 \times 10^{-6}$ & $6.64 \times 10^{-6}$ & $6.27 \times 10^{-6}$ & $6.22 \times 10^{-6}$ \\
$^{25}$Mg & $1.93 \times 10^{-4}$ & $2.69 \times 10^{-4}$ & $2.58 \times 10^{-4}$ & $2.49 \times 10^{-4}$ \\
$^{26}$Mg & $1.40 \times 10^{-5}$ & $1.68 \times 10^{-5}$ & $1.63 \times 10^{-5}$ & $1.60 \times 10^{-5}$ \\
$^{26}$Al & $5.33 \times 10^{-5}$ & $7.12 \times 10^{-5}$ & $6.86 \times 10^{-5}$ & $6.63 \times 10^{-5}$ \\
$^{27}$Al & $2.44 \times 10^{-4}$ & $2.95 \times 10^{-4}$ & $2.87 \times 10^{-4}$ & $2.80 \times 10^{-4}$ \\
	\midrule[0.3pt]\bottomrule[1pt] 
\end{tabular}
\label{table:CO2model}
\end{table*}

\begin{table*}[h!]
\centering
\caption{Predicted ejecta mass fractions for a 1.15 M$_{\odot}$ ONe nova model in the Ne-Al region. Total mass of the ejected envelope is $2.46 \times 10^{-5}$ M$_{\odot}$.}
\begin{tabular}{c c c c c}
\toprule[1pt]\midrule[0.3pt] \noalign{\smallskip}
	Nuclide & STARLIB-2013 & Low Rate & Medium Rate & High Rate \\ \cmidrule[0.3pt]{1-5} \noalign{\smallskip}
$^{20}$Ne & $1.76 \times 10^{-1}$ & $1.76 \times 10^{-1}$ & $1.76 \times 10^{-3}$ & $1.76 \times 10^{-1}$ \\
$^{21}$Ne & $3.89 \times 10^{-5}$ & $3.89 \times 10^{-5}$ & $3.89 \times 10^{-5}$ & $3.89 \times 10^{-5}$ \\
$^{22}$Ne & $6.51 \times 10^{-4}$ & $3.20 \times 10^{-4}$ & $3.58 \times 10^{-4}$ & $3.93 \times 10^{-4}$ \\
$^{22}$Na & $1.42 \times 10^{-4}$ & $1.42 \times 10^{-4}$ & $1.43 \times 10^{-4}$ & $1.42 \times 10^{-4}$ \\
$^{23}$Na & $1.01 \times 10^{-3}$ & $1.01 \times 10^{-3}$ & $1.04 \times 10^{-3}$ & $1.00 \times 10^{-3}$ \\
$^{24}$Mg & $1.44 \times 10^{-4}$ & $1.42 \times 10^{-4}$ & $1.52 \times 10^{-4}$ & $1.42 \times 10^{-4}$ \\
$^{25}$Mg & $3.52 \times 10^{-3}$ & $3.57 \times 10^{-3}$ & $3.56 \times 10^{-3}$ & $3.54 \times 10^{-3}$ \\
$^{26}$Mg & $2.98 \times 10^{-4}$ & $3.01 \times 10^{-4}$ & $3.04 \times 10^{-4}$ & $2.98 \times 10^{-4}$ \\
$^{26}$Al & $9.94 \times 10^{-4}$ & $1.01 \times 10^{-3}$ & $9.98 \times 10^{-4}$ & $1.01 \times 10^{-3}$ \\
$^{27}$Al & $8.54 \times 10^{-3}$ & $8.63 \times 10^{-3}$ & $8.59 \times 10^{-3}$ & $8.62 \times 10^{-3}$ \\
    \midrule[0.3pt]\bottomrule[1pt] 
\end{tabular}
\label{table:ONe1model}
\end{table*}

\begin{table*}[h!]
\centering
\caption{Predicted ejecta mass fractions for a 1.25 M$_{\odot}$ ONe nova model in the Ne-Al region. Total mass of the ejected envelope is $1.89 \times 10^{-5}$ M$_{\odot}$.}
\begin{tabular}{c c c c c}
\toprule[1pt]\midrule[0.3pt] \noalign{\smallskip}
	Nuclide & STARLIB-2013 & Low Rate & Medium Rate & High Rate \\ \cmidrule[0.3pt]{1-5} \noalign{\smallskip}
$^{20}$Ne & $1.78 \times 10^{-1}$ & $1.79 \times 10^{-1}$ & $1.79 \times 10^{-3}$ & $1.79 \times 10^{-1}$ \\
$^{21}$Ne & $3.64 \times 10^{-5}$ & $3.64 \times 10^{-5}$ & $3.64 \times 10^{-5}$ & $3.64 \times 10^{-5}$ \\
$^{22}$Ne & $1.30 \times 10^{-3}$ & $7.53 \times 10^{-4}$ & $8.23 \times 10^{-4}$ & $8.89 \times 10^{-4}$ \\
$^{22}$Na & $1.74 \times 10^{-4}$ & $1.74 \times 10^{-4}$ & $1.75 \times 10^{-4}$ & $1.74 \times 10^{-4}$ \\
$^{23}$Na & $1.11 \times 10^{-3}$ & $1.13 \times 10^{-3}$ & $1.15 \times 10^{-3}$ & $1.12 \times 10^{-3}$ \\
$^{24}$Mg & $1.08 \times 10^{-4}$ & $1.09 \times 10^{-4}$ & $1.24 \times 10^{-4}$ & $1.13 \times 10^{-4}$ \\
$^{25}$Mg & $2.27 \times 10^{-3}$ & $2.33 \times 10^{-3}$ & $2.32 \times 10^{-3}$ & $2.30 \times 10^{-3}$ \\
$^{26}$Mg & $1.67 \times 10^{-4}$ & $1.74 \times 10^{-4}$ & $1.77 \times 10^{-4}$ & $1.71 \times 10^{-4}$ \\
$^{26}$Al & $5.76 \times 10^{-4}$ & $5.76 \times 10^{-3}$ & $5.71 \times 10^{-4}$ & $5.77 \times 10^{-3}$ \\
$^{27}$Al & $4.53 \times 10^{-3}$ & $4.51 \times 10^{-3}$ & $4.50 \times 10^{-3}$ & $4.52 \times 10^{-3}$ \\
   \midrule[0.3pt]\bottomrule[1pt] 
\end{tabular}
\label{table:ONe2model}
\end{table*}

\end{document}